\documentclass[12pt]{article}

\newcommand{\blind}{0}

\input{myCommands.sty}

\begin{document}
		
	\def\spacingset#1{\renewcommand{\baselinestretch}%
		{#1}\small\normalsize} \spacingset{1}

	
	\if0\blind
	{
		\title{\bf Estimation and selection for high-order Markov chains with Bayesian mixture transition distribution models}

		\author[$\dag$]{Matthew Heiner \thanks{This research was supported in part by the National Science Foundation under award SES 1631963. The authors gratefully acknowledge contributions from Stephan Munch.}}
		\author[$\dag$]{Athanasios Kottas}
		\affil[$\dag$]{Department of Statistics, University of California, Santa Cruz, California, USA}

		\maketitle
	} \fi
	
	\if1\blind
	{
		\bigskip
		\bigskip
		\bigskip
		\begin{center}
			{\LARGE\bf Estimation and selection for high-order Markov chains with Bayesian mixture transition distribution models}
		\end{center}
		\medskip
	} \fi
	
	\bigskip
	\begin{abstract}
		We develop two models for Bayesian estimation and selection in high-order,\\ discrete-state Markov chains. Both are based on the mixture transition distribution, which constructs a transition probability tensor with additive mixing of probabilities from first-order transition matrices. We demonstrate two uses for the proposed models: parsimonious approximation of high-order dynamics by mixing lower-order transition models, and order/lag selection through over-specification and shrinkage via priors for sparse probability vectors. The priors further shrink all models to an identifiable and interpretable parameterization, useful for data analysis. We discuss properties of the models and demonstrate their utility with simulation studies. We further apply the methodology to a data analysis from the high-order Markov chain literature and to a time series of pink salmon abundance in a creek in Alaska, U.S.A.
	\end{abstract}
	
	\noindent%
	\vfill
	
	\newpage
	\spacingset{1.5} 

\section{Introduction}
  \label{sec:intro}
  
  Consider modeling a time series of nominal or ordinal values $s_t \in \{1, \ldots, K\}$ collected at equally spaced, discrete times $t=1,\ldots,T$. A popular approach for capturing serial correlation is to assume Markovian dynamics: that the conditional probability distribution of $s_t$ depends only on the recent past. Time homogeneity, or time invariance of the transition probabilities, is also typically assumed. These simplifying assumptions, nearly essential for inference in small or moderate sample size scenarios, are often appropriate even if the time series is not truly Markovian. Another common assumption is to condition only on the single most recent lag. However, restricting a model to first-order dynamics, or even selecting the incorrect lag, can miss important features in the data. 
  In this article, we propose Bayesian models to address two distinct objectives: estimation for the relevant time-delay coordinates, the Markovian order and important lags; and parsimonious modeling of high-order chains.
  
  Assuming time homogeneity, a full, unrestricted first-order model requires estimation of $K$ discrete distributions, each with $K-1$ free parameters. A Markov chain of order $L$ requires estimation of $K^L$ such distributions, limiting consideration to low orders for most time series. Typically, order (or lag) is selected by maximizing a (possibly penalized) likelihood \citep{katz1981mcBIC, raftery1985, prado2010}, performing trans-dimensional MCMC \citep{green1995reversible, insua2012}, using Bayes factors \citep{fan1999bayesFactMC, bacallado2011, zucchini2009hidden}, predictive criteria, or goodness-of-fit tests \citep{bartlett1951mc, besag2013exact}. Each of these approaches requires either fitting multiple models or complex estimation methods. Our approach is to build lag inference into a single model.
  
  Several approaches have been proposed to address exponential growth in the parameter space for higher-order transitions. \cite{raftery1985}
  introduced the mixture transition distribution (MTD), a general-purpose, parsimonious model for Markov chains. The MTD model was extended in \citet{raftery1994tavare} and developed over the subsequent decade. \citet{berchtold2002} provide a review. In the original MTD model, lags contribute to the transition probabilities by mixing over a single transition matrix. Only one new parameter is added for each additional lag. Despite its simplicity, the MTD framework can provide flexibility to capture nonstandard features, such as ``outliers, bursts, and flat stretches,'' as demonstrated by \citet{le1996gmtd} for a continuous-state version of the MTD.
  
  Contemporary with the MTD model, generalized linear models for multinomial outcomes were applied to categorical time series \citep{liang1986, zeger1986, fahrmeir1987}. These models can accommodate varying degrees of complexity by controlling the order of interactions among the linear predictors (lags), up to and including a full model with $K^L(K-1)$ parameters. These models can also account for exogenous sources of non-stationarity through covariates. However, estimation and interpretability become problematic in these models when many lags are considered.
  
  Tree-based methods provide an alternative parsimonious approach. Variable-length Markov chains (VLMC, \citealp{ron1994, buhlmann1999}) reduce the parameter space by clustering the $K^L$ transition distributions via recursive pruning. Sparse Markov chains (SMC, \citealp{jaaskinen2014}) partition the $L$-dimensional lag space without hierarchical constraints, resulting in greater flexibility. They also feature a prior structure which encourages low orders. Although efficient, these models lack posterior uncertainty quantification, and inferences for order and lag importance are not readily available.
  
  More recently, \citet{sarkar2016} proposed a Bayesian nonparametric model for high-order Markov chains. They model the $K^L$ transition distributions through tensor factorization and further encourage parsimony by clustering the components of a core mixing distribution with a Dirichlet process prior \citep{ferguson1973}. By allowing variable dimensions along different modes of the core mixing distribution, the model further admits inferences for lag importance. This model enjoys a fully Bayesian, albeit complicated, implementation and performs well against the methods described above in forecasting when there are up to four states and ten lags.
  
  Our modeling strategy is to build on the simplicity and interpretability of the MTD model. One popular extension of the MTD, referred to by \citet{berchtold2002} as the MTDg model, utilizes a separate transition matrix for each lag. While this more flexible model grows linearly with each additional lag, it is not identifiable \citep{lebre2008mtd_em}. Recently, \citet{tank2017arXiv} used a reparameterization to establish a unique and identifiable characterization of the MTDg model in the context of multiple time series. Using a penalized likelihood and proximal gradient optimization, they softly enforce the identifiability conditions and simultaneously select relevant series to infer Granger causality. We propose a Bayesian estimation approach to the MTDg model which utilizes priors introduced by \citet{heiner2019spv} to promote shrinkage toward the identifiability conditions of \citet{tank2017arXiv}, and to simultaneously select relevant lags. These priors were previously demonstrated to effectively select a single active lag using the original MTD model. We then propose an extension which allows for higher-order interaction between lags, as well as inference for the Markovian order (i.e., the number of active lags) up to a pre-specified maximum. 
  
  The remainder of this paper is organized as follows. In Section \ref{sec:models}, we review the MTD model and develop our proposed extensions. We outline our approach for Bayesian inference using structured priors to aid with the models' intended uses in Section \ref{sec:inference}. In Section \ref{sec:sims}, we test the models using two simulation scenarios that reflect our two objectives, demonstrating improved predictive performance over the original MTD. Section \ref{sec:applications} illustrates the models through two analyses, first on a data set which appears in the preceding literature, and second on annual time series of pink salmon abundance in Alaska, U.S.A. Finally, we conclude with a summary in Section \ref{sec:discussion}. Technical details are provided in the appendices.

\section{Models}
  \label{sec:models}

In a full $L$-order, time-homogeneous Markov chain, the collection of all possible transition probabilities $\Pr(s_t = k_0 \mid s_{t-1} = k_1, \ldots, s_{t-L} = k_L)$, for $k_\ell \in \{1, \ldots, K\}$, $\ell \in \{1, \ldots, L\}$, $t \in \{L+1, \ldots, T\}$, can be arranged in a $(L+1)$-order tensor $(\bm{\Omega})_{k_0, k_{1}, \ldots, k_{L}}$. If we condition on the first $L$ observations of the time series, the joint sampling distribution for the remaining sequence is given by $\Pr(\{s_t\}_{t=L+1}^T \mid \{s_t\}_{t=1}^L, \bm{\Omega}) = \prod_{t=L+1}^{T} (\bm{\Omega})_{s_{t}, s_{t-1}, \ldots, s_{t-L}}$, defining the conditional likelihood that we employ hereafter. We begin by specifying the original MTD model in Section \ref{sec:mtd} and motivate its extensions. In Section \ref{sec:mtdg}, we discuss the MTDg extension and associated identifiability results. We then introduce a Bayesian formulation for the MTDg which uses priors for sparse probability vectors in Section \ref{sec:spv}. Finally, we propose an extension to include higher-order transitions in Section \ref{sec:mmtd}.

 \subsection{Original mixture transition distribution}
  \label{sec:mtd} 
  
The mixture transition distribution model constructs the transition probability tensor $\bm{\Omega}$ as linear combinations of probabilities from a single column-stochastic matrix $\bm{Q}$ and adds just one parameter for each additional lag $(\lambda_\ell)$, similar to autoregressive models. The transition probabilities in a model of order $L$ are given as
\begin{align}
\label{eq:MTDtrans}
\Pr(s_t=k_0 \mid s_{t-1}=k_1, \ldots, s_{t-L}=k_L)  
= \sum_{\ell=1}^L \lambda_\ell \, q_{k_0, k_\ell}  \, ,
\end{align}
where $q_{i,j} \equiv (\bm{Q})_{i,j}$, $0 \le \lambda_\ell \le 1$ and $\sum_{\ell=1}^L \lambda_\ell = 1$. Although this model incorporates information beyond the first lag, it is restrictive in that it cannot capture nonlinear (non-additive) dynamics in more than one dimension of the lag space.

Form (\ref{eq:MTDtrans}) suggests that lags which play a prominent role in the transition probability for $s_t$ will have relatively large $\lambda_\ell$ and lags which are not important to the transition will have $\lambda_\ell$ values near 0. Hence, inferences for $\bm{\lambda}=(\lambda_1, \ldots, \lambda_L)$ potentially yield information about important lags for the Markov process. It is apparent from (\ref{eq:MTDtrans}) that $\lambda_\ell = 0$ is sufficient for conditional independence of $s_t$ and $s_{t-\ell}$. If the columns of $\bm{Q}$ are unique, then $\lambda_\ell = 0$ is also a necessary condition for conditional independence. Inferences on $\bm{\lambda}$ have been employed to understand lag importance informally \citep{raftery1994tavare}, although the standard method for assessing order has been to compare BIC values \citep{berchtold2002}. \citet{heiner2019spv} use a single model, relying on inferences on $\bm{\lambda}$ for insight into lag importance. We adopt the same approach here.

\subsection{MTDg, identifiability, and lag selection}
\label{sec:mtdg}

The MTDg model modifies (\ref{eq:MTDtrans}) by using a distinct column-stochastic matrix $\bm{Q}^{(\ell)}$ for each lag $\ell = 1, \ldots, L$. While this increases flexibility and allows for different transition types associated with each lag, the model lacks identifiability. \citet{tank2017arXiv} demonstrate this by introducing an intercept probability vector, $\bm{Q}^{(0)} = \left(q_1^{(0)}, \ldots, q_K^{(0)} \right)$, extending $\bm{\lambda}$ to include $\lambda_0$, and reparameterizing the transition probabilities through the products $\varphi^{(0)}_{k_0} \equiv \lambda_0 \, q^{(0)}_{k_0}$ and $\varphi^{(\ell)}_{k_0, k_\ell} \equiv \lambda_\ell \, q^{(\ell)}_{k_0, k_\ell}$, resulting in the MTDg formulation
\begin{align}
\label{eq:MTDgTank}
\Pr(s_t = k_0 \mid s_{t-1} = k_1, \ldots, s_{t-L} = k_L) = \varphi_{k_0}^{(0)} + \sum_{\ell=1}^L \varphi^{(\ell)}_{k_0, k_\ell} \, .
\end{align}
One can then freely transfer probability mass by subtracting some vector \\ 
$\bm{a}_\ell = \left( a_1^{(\ell)}, \ldots, a_K^{(\ell)} \right)$ from each column of $\bm{\varphi}^{(\ell)} \equiv \lambda_\ell \, \bm{Q}^{(\ell)}$ and adding it to $\bm{\varphi}^{(0)}$ while preserving all values in $\bm{\Omega}$. Selecting $a_k^{(\ell)}$ to be the minimum value in the $k$th row of $\bm{\varphi}^{(\ell)}$ for each $k=1,\ldots,K$, and following the transferral procedure just described for $\ell=1,\ldots,L$, results in a maximally reduced parameterization in the sense that the highest probability mass possible has been transferred to the intercept while maintaining non-negativity of all elements in $\bm{\varphi}^{(\ell)}$. Let $\{\tilde{\bm{\varphi}}^{(\ell)}\}_{\ell=0}^L$ denote the resulting parameters after the reduction procedure so that $\tilde{\bm{\varphi}}^{(0)} 
\equiv \bm{\varphi}^{(0)} + \sum_{\ell = 1}^L \bm{a}_\ell $ and $ \tilde{\varphi}^{(\ell)}_{i,j} \equiv \varphi^{(\ell)}_{i,j} - a^{(\ell)}_i$, for $i = 1, \ldots, K$, $j = 1, \ldots, K$, and $\ell = 1, \dots, L$. \citet{tank2017arXiv} show that this maximal reduction yields a unique representation for every MTDg. Furthermore, one can view the reduced model in the original parameterization using $\tilde{\bm{\lambda}} = (\tilde\lambda_0, \ldots, \tilde\lambda_L)$ with $\tilde\lambda_0 \equiv \sum_{k=1}^K \tilde{\varphi}^{(0)}_k $, and $\tilde\lambda_\ell \equiv \sum_{k=1}^K \tilde{\varphi}^{(\ell)}_{k,j} = \lambda_\ell - \sum_{k=1}^K a^{(\ell)}_k $, for $\ell = 1, \ldots, L$, and invariant to choice of $j$; probability vector $\tilde{\bm{Q}}^{(0)} \equiv (\tilde{\lambda}_0)^{-1} \,  \tilde{\bm{\varphi}}^{(0)} $; and column-stochastic matrices $\tilde{\bm{Q}}^{(\ell)} \equiv (\tilde{\lambda}_\ell)^{-1} \, \tilde{\bm{\varphi}}^{(\ell)}$. Then $\tilde{\lambda}_\ell$ is interpretable as a marginal contribution of the $\ell$th lag to the transition distribution. Thus, with careful construction, we may use $\tilde{\bm{\lambda}}$ to make inferences about lag importance even though the MTDg is overparameterized. Furthermore, the intercept allows us to infer the possible lack of serial dependence in a direct way.

To operationalize this reduction in an estimation procedure, \citet{tank2017arXiv} show that solutions to a penalized likelihood based on (\ref{eq:MTDgTank}), in which the penalty increases with respect to the absolute value of the entries in the $\{\bm{\varphi}^{(\ell)}\}_{\ell=1}^L$ (excluding the intercept), meet the maximal-reduction criterion. Their proposed soft penalty functions equivalently regularize $\lambda_\ell$ for $\ell = 1, \ldots, L$. Because $\lambda_\ell = 0$ is sufficient and necessary (as long as the columns of $\bm{Q}^{(\ell)}$ are distinct) for conditional independence of the current state from lag $\ell$, the penalized estimate simultaneously detects lag relevance (or Granger causality in the case where $\ell$ indexes multiple time series).

\subsection{Bayesian MTDg with priors for sparse probability vectors}
 \label{sec:spv}

We now present a Bayesian modeling approach to the MTDg, which admits full characterization of uncertainty. Rather than addressing the constraint that each column of $\bm{\varphi}^{(\ell)}$ sum to $\lambda_\ell$, we work with the original $\bm{\lambda}$ and $\{\bm{Q}^{(\ell)}\}$ parameters and employ carefully chosen prior distributions that shrink toward the identifiable and interpretable $\tilde{\bm{\lambda}}$ and $\{\tilde{\bm{Q}}^{(\ell)}\}$. In this and the following section, we employ two prior distributions for probability vectors, introduced by \citet{heiner2019spv}, which go beyond the standard Dirichlet prior by enforcing sparsity in the presence of data, as well as conditional stochastic ordering. Both priors are continuous, bypassing problems that arise from the sum-to-one constraint when using priors with point masses.

The first is the sparse Dirichlet mixture (SDM) prior. This one-parameter extension of the Dirichlet distribution is a fixed-weight mixture of Dirichlet distributions, with each component featuring a boost of equivalent sample size $\beta > 1$ in one of the categories. If $\bm{\theta}$ is a probability vector of length $J$, the SDM density is given as
 \begin{align}
\label{eq:SDMpdf}
p_{\SDM}(\bm{\theta}; \bm{\alpha}, \beta)  =  \sum_{j=1}^J \frac{w_j}{ \sum_{i=1}^J w_i } \Dirdist(\bm{\theta}; \bm{\alpha} + \beta \bm{\mathrm{e}}_j ) \, ,
\end{align}
where $\Dirdist(\bm{\theta}; \bm{\alpha} + \beta {\bm{\mathrm{e}}}_j )$ denotes the Dirichlet density with shape parameter vector $\bm{\alpha} + \beta {\text{\bf e}}_j$, with $\bm{\alpha} = (\alpha_1, \ldots, \alpha_J)$, where $\bm{\mathrm{e}}_j$ is a vector of 0s with a 1 in the $j$th position, and $w_j \equiv \prod_{h=1}^J \Gamma(\alpha_h + \beta 1_{(h=j)})$ with $\Gamma(\cdot)$ denoting the gamma function. For small sample sizes and relatively large $\beta$, the SDM can be described as a winner-takes-all prior in that it shifts most mass toward the $\theta_j = 1$ corner of the supporting simplex for the component with largest $\alpha_j$.

The second prior is the stick-breaking mixture (SBM) prior. This prior builds the probability vector $\bm{\theta}$ through an extension of the stick-breaking construction that defines the generalized Dirichlet distribution \citep{connor1969}. In particular,
\begin{align}
\label{eq:stickbreaking}
\theta_1 = X_1, \ \theta_j = X_j \prod_{i=1}^{j-1} (1 - X_i) \ \text{for}\ j=2,\ldots,J-1, \ \text{and} \ \theta_J = \prod_{i=1}^{J-1} (1 - X_i) \, ,
\end{align}
with $X_j$ independently drawn from a mixture of three beta distributions, $X_j \simindep \\ 
\pi_1 \Betadist(1, \eta) + \pi_2 \Betadist(\gamma_j, \delta_j) + \pi_3 \Betadist(\eta, 1)$, where $\pi_1 + \pi_2 + \pi_3 = 1$. This mixture structure encourages sparsity by setting $\eta$ large, in which case the first component corresponds to small probabilities in $\bm{\theta}$. The third component allows for the rest of the unbroken stick (i.e., $\prod_{i=1}^{j-1}(1-X_i) = 1 - \sum_{i=1}^{j-1} \theta_i$) to be used for $\theta_j$, while the second mixture component allows for flexibility in modeling $\theta_j$. The $\gamma_j$ and $\delta_j$ parameters can be fixed at the same values across $j$, or can be set to mimic the Dirichlet distribution as with the generalized Dirichlet distribution, resulting in a three-parameter extension of the Dirichlet distribution. To accommodate our proposed extensions of the MTD model, we adapt the SBM prior to allow the mixture weights $\pi_1$, $\pi_2$, and $\pi_3$ to vary with $j$.

Beyond the properties noted, one important advantage of the SDM and SBM priors is their conjugacy and resulting computational tractability. If the hyperparameters of the priors are fixed, as is usually the case with the original Dirichlet priors, incorporating them into a hierarchical model involving multinomial counts (latent or observed) requires minimal effort since posterior Gibbs sampling proceeds with conditional distributions that can be directly sampled, allowing us to swap priors without structural changes to the updates in Appendix \ref{sec:appendix_mtdg}.

If a modeler believes that exactly one lag influences the transition distribution, the SDM prior can be used in the single-$\bm{Q}$ MTD model, as demonstrated in \citet{heiner2019spv}. However, this is not recommended for the MTDg model. If $\beta$ is not sufficiently high, the SDM prior may distribute posterior mass to a non-unique and non-interpretable configuration of $\bm{\lambda}$. We instead use a SBM prior for $\bm{\lambda}$ that favors the unique reduction and more appropriately allows for dependence on multiple lags. Here, the stick-breaking construction of the SBM provides intuition, as $\lambda_0$ is drawn first, and the rest of $\bm{\lambda}$ is broken sequentially from what remains in the unbroken stick. 
To avoid penalizing the intercept, we set $\pi_2 = 1$ for $\lambda_0$ only and use either $\gamma_0 = \delta_0 = 1$ (the uniform distribution) or beta shape parameters that favor large values of $\lambda_0$. The remaining beta mixtures use $\pi_1 > 0$ and $\pi_3 > 0$ to regularize $\{\lambda_\ell\}_{\ell=1}^L$. Setting $\pi_1 > 0$ allows for small values of the corresponding $\lambda_\ell$, effectively skipping the $\ell$th lag. Setting $\pi_3 > 0$ promotes consumption of the remaining mass before reaching $\lambda_L$. If $\gamma_\ell = \gamma$ and $\delta_\ell = \delta$ across $\ell$ and $\pi_2$ is relatively high, the sequential SBM construction can further regularize $\bm{\lambda}$ via stochastic ordering, consistent with the common assumption that recent lags should carry more influence.

The model is completed with prior distributions for $\{ \bm{Q}^{(\ell)} \}$. The traditional choice for transition matrices is to use independent Dirichlet distributions for each column, which we adopt here. \citet{heiner2019spv} found it advantageous to use independent SBM priors for each column of $\bm{Q}$ (in the standard MTD model) in cases of nearly deterministic dynamics. However, the MTDg model spreads estimation across multiple $\bm{Q}^{(\ell)}$ matrices, relying more heavily on the non-symmetric SBM prior. This can potentially introduce undesired artifacts to the estimated transition probabilities.

The full hierarchical model specification for the MTDg and details for posterior inference are discussed in Section \ref{sec:inference} and Appendix \ref{sec:appendix_mtdg}.

\subsection{Mixtures of higher-order MTD components}
 \label{sec:mmtd}

The MTD and MTDg models offer parsimonious and interpretable representations for Markov chains with dependence extending beyond the most recent lag. However, these models are strictly additive in the sense that any dynamics of order higher than one (i.e., more than one active lag) are approximated with linear combinations of first-order transitions. In their survey of generalizations for the MTD, \citet{berchtold2002} suggest, but do not pursue, the possibility of mixing over higher-order transition tensors. 
We build a Bayesian framework for such an extension to include higher-order ``interactions,'' and we refer to the resulting model as the mixture of mixture transition distributions (MMTD) model.

To define the MMTD model, let $R < L$ be a positive integer representing the highest-order transition tensor over which we will mix. Thus we have $\bm{\mathcal{Q}}^{(0)}$, a length-$K$ probability vector; $\bm{\mathcal{Q}}^{(1)}$, a $K \times K$ transition matrix; $\bm{\mathcal{Q}}^{(2)}$, a $K \times K \times K$ transition tensor; and so forth up to $\bm{\mathcal{Q}}^{(R)}$, a $K^{R+1}$ transition tensor, such that $\sum_{k=1}^K (\bm{\mathcal{Q}}^{(R)})_{k, k_1, \ldots, k_R} = 1$ for all $({k_1, k_{2}, \ldots, k_R}) \in \{1, \ldots, K \}^R$. Next, introduce a mixing probability vector across orders, $\bm{\Lambda} = (\Lambda_0, \Lambda_1, \ldots, \Lambda_R)$. The MMTD model for transition probabilities is then given by
\begin{align}
\label{eq:MMTDtrans}
\Pr(s_t &=k_0 \mid s_{t-1}=k_1, \ldots, s_{t-L}=k_L) = (\bm{\Omega})_{k_0, k_{1}, \ldots, k_L} \nonumber \\
&= \Lambda_0 \, (\bm{\mathcal{Q}}^{(0)})_{k_0} + \Lambda_1 \sum_{\ell=1}^L \lambda_{\ell}^{(1)} \, (\bm{\mathcal{Q}}^{(1)})_{k_0, k_\ell} \, + \nonumber \\
& \qquad \qquad \quad \ \ \, \,  + \Lambda_2  \underset{1 \le \ell_1 < \ell_2 \le L}{\sum \sum} \lambda_{(\ell_1, \ell_2)}^{(2)} \, (\bm{\mathcal{Q}}^{(2)})_{k_0, k_{\ell_1}, k_{\ell_2}} \, + \ldots \\
& \qquad \qquad  \ \ldots\, \,  + \Lambda_R \underset{1 \le \ell_1 < \ldots < \ell_R \le L}{\sum \ldots \sum} \lambda_{(\ell_1, \ldots, \ell_R)}^{(R)} \,  (\bm{\mathcal{Q}}^{(R)})_{k_0, k_{\ell_1}, \ldots, k_{\ell_R} } \nonumber \, ,
\end{align}
where $\bm{\lambda}^{(r)}$ is a probability vector of length $L \choose r$ for $r = 1, \ldots, R$. This mixture of mixtures is equivalent to using a single (albeit long) $\bm{\lambda}$ probability vector to mix over all possible arrangements of lags and base transition tensors $\bm{\mathcal{Q}}^{(r)}$. However, this parameterization is more informative about important orders (via inference for $\bm{\Lambda}$) in addition to lags (via inference for $\bm{\lambda}^{(r)}$). If $\Lambda_1 = 1$, we recover the original MTD model. The fully-parameterized transitions associated with $\bm{\mathcal{Q}}^{(r)}$ allow unrestricted dynamics in $r$ dimensions of the lag space. As a discrete mixture of probability distributions, this model produces a valid probability tensor.

The model in (\ref{eq:MMTDtrans}) is clearly over-parameterized, and consequently $\bm{\Lambda}$, $\{  \bm{\lambda}^{(r)} \}_{r=1}^R$, and $\{\bm{\mathcal{Q}}^{(r)}\}_{r=0}^R$ are not fully identified. Defining a reduction procedure similar to that of \citet{tank2017arXiv} for the MMTD is more nuanced. One complication arises because the result is dependent on the order of reduction. For example, one may first transfer probability mass to the intercept from all higher-order transition tensors, followed by transfer to the first-order transitions by defining elements of the $\bm{a}$ vector as the minima over indexes in $\bm{\mathcal{Q}}^{(r)}$, for $r \ge 2$, which correspond to a unique value of the lagged state for the lag associated with the current $\bm{\mathcal{Q}}^{(1)}$ (allowing for $L$ such matrices). However, transferring first to the $\bm{\mathcal{Q}}^{(1)}$ associated with lag 1 and then to the $\bm{\mathcal{Q}}^{(1)}$ associated with lag 2 does not yield the same result if we reverse the order. Alternatively, one could define a reduction process in terms of projecting $\bm{\Omega}$ first onto an intercept, then projecting what remains onto the space spanned by the first-order level of the MMTD, and so forth. Thus, the intercept has the first opportunity to describe the base probabilities, then the first-order level of the model has the next opportunity to capture first-order dynamics, and each additional level fills in what lower-order levels cannot adequately model. Absent a formal procedure, we note that in estimation, this process would be implemented with regularization, for which the sequential SBM prior is well-suited. We therefore propose using a SBM prior for $\bm{\Lambda}$, similar to the one used in Section \ref{sec:spv} for the MTDg model.

Even under regularization of model order, it is possible 
for a high-order $\bm{\mathcal{Q}}^{(r)}$ to mimic a lower-order tensor through repetition of transition probabilities across values of a certain lag. Rather than build complicated constraints into the model, we note that this issue can be detectable through inferences for $\{\bm{\mathcal{Q}}^{(r)}\}_{r=1}^R$. Specifically, a modeler may plot posterior estimates of the distributions in $\{\bm{\mathcal{Q}}^{(r)}\}_{r=1}^R$ and check for repeating patterns, especially patterns that coincide with a slice of the tensor (e.g., all columns equal in $\bm{\mathcal{Q}}^{(1)}$ in a $R=1$ model would indicate that the intercept alone is adequate). We strongly recommend following this practice before interpreting model inferences for $\bm{\Lambda}$ and $\{ \bm{\lambda}^{(r)} \}_{r=1}^R$.

We envision two primary uses for the MMTD model. The first is to uncover low-order structure from data whose practical lag dependence horizon is truly smaller than our over-specified $L$ and the order is less than or equal to our selected $R$, in which case the true model is contained within the mixture framework. For example, we might postulate that a time series has second-order dependence, but we are unsure which two lags are important. Assuming a maximal lag horizon of 10, we could fit the MMTD model with $L=10$ and $R \ge 2$. Because there is only one $\bm{\mathcal{Q}}^{(r)}$ at each level, we could use the SDM prior on each $\bm{\lambda}^{(r)}$ with a large value of $\beta$ to select the appropriate lag configuration and discourage mixing lower-order transitions. If the dynamics are truly second-order, we would anticipate that $\Lambda_2$ would carry substantial posterior weight, and that inferences for $\bm{\lambda}^{(2)}$ would identify the influential lags. In this model-selection scenario, the SDM (on $\bm{\lambda}$) and SBM (on $\bm{\Lambda}$) priors play an important role in selection and interpretation.

If the true order of dependence in the time series is greater than $R$, our second intended use for the MMTD model is analogous to that of the MTD and MTDg, wherein we parsimoniously approximate higher-order dependence by mixing lower-order transition distributions. Adding the higher-order $\bm{\mathcal{Q}}^{(r)}$ tensors could be thought of as including interaction-like terms in the mixture. In this scenario, one may still use the SDM prior for each $\bm{\lambda}^{(r)}$, but with a lower value of $\beta$ to encourage more mixing (note that $\beta=1$ yields a Dirichlet prior). The SBM prior on $\bm{\Lambda}$ further allows mixing across orders, so that different levels of the model may mix across non-overlapping sets of lags.

As with the MTDg, we recommend using independent Dirichlet priors for each of the $K^r$ probability distributions in $\bm{\mathcal{Q}}^{(r)}$ for $r=0,1,\ldots,R$. If $L$ and $R$ are small, $T$ is large, and the transition probability tensor is known to be sparse, it {\it may potentially} be advantageous to replace these Dirichlet priors with independent SBM priors. However, we strongly urge caution, as information from the data is spread thin and the SBM prior is not symmetric. \citet{heiner2019spv} discuss a strategy for promoting more Dirichlet-like behavior in the SBM prior.

Our proposed model formulation requires estimation of $R$ free $\Lambda$ parameters, ${L \choose 1} + {L \choose 2} + \ldots + {L \choose R} - R$ free $\lambda$ parameters, and $ (K-1) + K(K-1) + K^2(K-1) + \ldots + K^R(K-1) = K^{R+1} - 1$ free parameters in $\{\bm{\mathcal{Q}}^{(r)}\}_{r=0}^R$. The fastest-growing term in the $\lambda$ parameter count increases no faster than a polynomial in $L$ of degree ${\lfloor R/2 \rfloor}$ divided by $R!$, while the transition distributions grow exponentially. Table \ref{tab:MMTDparamCount} reports the total number of parameters to estimate for different combinations of $K$, $L$, and $R$. Typically, $K$ is fixed and known, and a modeler must select $L$ and $R$ considering parsimony, estimability for a given sample size, and computational cost. If $R$ is much smaller than $L$, the MMTD substantially reduces the parameter space from the original full-order Markov chain. The parameter space is effectively further reduced by the sparsity-inducing priors on $\bm{\Lambda}$ and $\{\bm{\lambda}^{(r)}\}$.

The hierarchical model specification for the MMTD and posterior inference details are discussed in Section \ref{sec:inference} and Appendix \ref{sec:appendix_mmtd}.

\begin{table}[t]
	\caption{Free parameter count for MMTD model under different combinations of state-space size $K$, largest possible lag $L$, and largest mixing order $R$. The total number of parameters is the sum of the free $\Lambda$, $\lambda$, and $\mathcal{Q}$ parameters. The unrestricted total is the number of parameters required to estimate an unrestricted transition probability tensor of order $L$.}
	\label{tab:MMTDparamCount}
	
	\begin{tabular}{ccccrrrcrr}
		\toprule
		$K$ & $L$ & $R$ & & $\Lambda$ &  $\lambda$ & $\mathcal{Q}$ & & Total & Unrestricted \\
		\midrule
		2 & 5 & 2 & & 2 & 13 & 7 & & 22 & 32		\\
		2 & 5 & 4 & & 4 & 26 & 31 & & 61 & 32		\\
		2 & 10 & 2 & & 2 & 53 & 7 & & 62 & 1,024	\\
		2 & 10 & 4 & & 4 & 381 & 31 & & 416 & 1,024	\\
		5 & 5 & 2 & & 2 & 13 & 124 & & 139 & 12,500	\\
		5 & 5 & 4 & & 4 & 26 & 3,124 & & 3,154 & 12,500	\\
		5 & 10 & 2 & & 2 & 53 & 124 & & 179 & 3.91$\times 10^7$ \\
		5 & 10 & 4 & & 4 & 381 & 3,124 & & 3,509 & 3.91$\times 10^7$ \\
		7 & 5 & 2 & & 2 & 13 & 342 & & 357 & 100,842		\\
		7 & 5 & 4 & & 4 & 26 & 16,806 & & 16,836 & 100,842		\\
		7 & 10 & 2 & & 2 & 53 & 342 & & 397 & 1.69$\times 10^9$	\\
		7 & 10 & 4 & & 4 & 381 & 16,806 & & 17,191 & 1.69$\times 10^9$	\\
		\bottomrule
	\end{tabular}
\end{table}

\section{Bayesian inference and computation}
 \label{sec:inference}

We now address implementation of the MTDg 
and MMTD models
. To obtain full posterior inference, we utilize a Gibbs sampler which alternates between collapsed and full conditional distributions made tractable by augmentation with latent configuration variables \citep{insua2012}. As noted earlier, all inferences condition on the first $L$ observations in the time series $\{s_t\}_{t=1}^T$.

\subsection{MTDg model}
\label{sec:MTDg_inference}

The sampling distribution for the MTDg model is
\begin{align}
\label{eq:lik_mtdg}
p(\{ s_t \}_{t=L+1}^T \mid \bm{\lambda}, \{ \bm{Q}^{(\ell)} \}_{\ell=0}^L, \{s_{t}\}_{t=1}^L) = \prod_{t=L+1}^T \left[ \lambda_0 \, q_{s_t}^{(0)} + \sum_{\ell=1}^L \lambda_{\ell} \, q_{s_t, s_{t-\ell}}^{(\ell)}  \right] \, .
\end{align}
We first break the mixture in (\ref{eq:lik_mtdg}) by introducing latent indicators $z_t$ such that $\Pr(z_t=\ell \mid \bm{\lambda}) = \lambda_\ell$ for $\ell=0,1,\ldots,L$ independently across $t$. Adding the priors yields the full hierarchical model. For $t=L+1, \ldots, T$; $k = 1, \ldots, K$; $i = 1, \ldots, K$; and $\ell=0, \ldots, L$, we have
\begin{align}
\label{eq:hier_mtdg}
&\bm{Q}^{(0)} \sim \Dirdist(\bm{\alpha}^{(0)}), \quad (\bm{Q}^{(\ell)})_{\cdot,i} \simindep \Dirdist(\bm{\alpha}_{i}^{(\ell)}) \text{ for } \ell > 0, \quad \bm{\lambda} \sim \SBM(\bm{\pi}_1, \bm{\pi}_3, \eta, \bm{\gamma}, \bm{\delta} ), \nonumber \\
&\Pr( z_t = \ell \mid \bm{\lambda} ) = \lambda_\ell \, , \\
&\Pr( s_t = k \mid z_t = \ell, \{s_{t'}\}_{t'=t-L}^{t-1}, \bm{Q}^{(\ell)}) = q_{k}^{(0)} 1_{(\ell=0)} + q_{k, s_{t-\ell}}^{(\ell)} 1_{(\ell>0)} \, ,  \nonumber
\end{align}
where each $\bm{\alpha}$ is a length-$K$ vector of positive shape parameters; $\bm{\pi}_1$ and $\bm{\pi}_3$ are length-$L$ vectors containing probabilities such that $(\bm{\pi}_1)_{\ell} + (\bm{\pi}_3)_{\ell} < 1$; and $\bm{\gamma}$ and $\bm{\delta}$ are length-$L$ vectors containing positive shape parameters. We always set $(\bm{\pi}_1)_{0} = 0$ to avoid penalizing $\lambda_0$, and also recommend setting $(\bm{\pi}_3)_{0} = 0$.

This structure admits closed-form full conditional distributions for all of $\{ z_t \}$, $\{ \bm{Q}^{(\ell)} \}$, and $\bm{\lambda}$. Specifically, the update for $\bm{\lambda}$ is a conjugate SBM-multinomial update using aggregated counts of $\{ z_t \}$. Each $z_t$ can be updated with a discrete distribution involving $\bm{\lambda}$ and elements of $\{ \bm{Q}^{(\ell)} \}$ as they appear in the likelihood. Given $\{ z_t \}$, we can aggregate the transition counts into sufficient statistics $\bm{N}^{(0)} = (n_1^{(0)}, \ldots, n_K^{(0)})$ and $\{\bm{N}^{(\ell)}\}_{\ell=1}^L$, a set of $K \times K$ matrices. For example, if $s_t = 1$, $z_t = 2$, and $s_{t-2} = 3$, we would increment $(\bm{N}^{(2)})_{1,3}$. The full conditional distribution for the intercept $\bm{Q}^{(0)}$ is then an updated Dirichlet distribution with $\bm{N}^{(0)}$ providing the multinomial counts. Likewise, the update for $(\bm{Q}^{(\ell)})_{\cdot,i}$ involves a conjugate Dirichlet-multinomial update with its corresponding count vector $(\bm{N}^{(\ell)})_{\cdot,i}$, for each $\ell = 1, \ldots, L$ and $i=1, \ldots, K$.

As is common with mixture models, the full joint posterior distribution is multimodal and the Gibbs sampler described above is prone to poor mixing. To improve mixing, we modify the Gibbs sampler just described in two ways. First, we integrate all $\{ \bm{Q}^{(\ell)} \}_{\ell=0}^L$ parameters out of the full joint posterior. This affects only the full conditional distributions for the configuration variables $\{ z_t \}$, which are drawn from a (different) discrete distribution. The second modification is an occasional (every 10 iterations) hybrid Metropolis step that jointly proposes $\bm{\lambda}$ and $\{ z_t \}$ from the prior in order to encourage exploration. Ordinarily, the prior is inefficient as a proposal distribution. While the sparse configurations proposed by the SBM prior help mitigate this issue, we still advocate running multiple long MCMC chains to ensure adequate mixing. Full details for the modified Gibbs sampler are provided in Appendix \ref{sec:appendix_mtdg}.



\subsection{MMTD model}
\label{sec:MMTD_inference}

Our implementation for the MMTD model is analogous to the MTDg model, with a few notable extensions. As before, the sampling distribution for the time series is given as a product of transition probabilities in (\ref{eq:MMTDtrans}) across $t=L+1, \ldots, T$. We again break the mixture in (\ref{eq:MMTDtrans}) by introducing latent configuration variables $Z_t$ such that $\Pr(Z_t = r \mid \bm{\Lambda} ) = \Lambda_r$, for $r = 0, 1, \ldots, R$, independently for each observation time. Then conditional on $Z_t$ (and for $Z_t > 0$), further introduce $\bm{z}_t$ such that $\Pr(\bm{z}_t = (\ell_1, \ldots, \ell_r) \mid Z_t = r, \bm{\lambda}^{(r)} ) = \lambda_{(\ell_1 , \ldots, \ell_r )}^{(r)}$, for $1 \le \ell_1 < \ldots < \ell_r \le L$, independently for each observation time. The hierarchical formulation for this model is given in generative order as follows. For $t = L+1, \ldots, T$; $k = 1, \ldots, K$; $k_\ell = 1, \ldots, K$; $\ell = 1, \ldots, L$, $1 \le \ell_1 < \ldots < \ell_r \le L$; and $r=0,1, \ldots, R$, we have
\begin{align}
\label{eq:MMTDhier}
&\bm{\mathcal{Q}}^{(0)} \sim \Dirdist(\bm{\alpha}_{Q^{(0)}}), \quad (\bm{\mathcal{Q}}^{(r)})_{\cdot, k_1, \ldots, k_r} \simindep \Dirdist(\bm{\alpha}_{Q^{(r)}}), \ \text{for} \ (k_1, \ldots, k_r) \in \{1, \ldots, K\}^r \, , \nonumber \\
&\bm{\Lambda} \sim \SBM(\bm{\pi}_1, \bm{\pi}_3, \eta, \bm{\gamma}, \bm{\delta} ) \, ,  \quad \quad 
\bm{\lambda}^{(r)} \simindep \SDM(\bm{\alpha}_{\lambda^{(r)}}, \beta_{\lambda^{(r)}}), \ \text{for} \ r > 0 \, , \nonumber \\
&\Pr(Z_t = r \mid \bm{\Lambda}) = \Lambda_r \, , \qquad 
\Pr(\bm{z}_t = (\ell_1, \ldots, \ell_r) \mid Z_t = r, \bm{\lambda}^{(r)}) = \lambda_{(\ell_1 , \ldots, \ell_r)}^{(r)} \, , \nonumber \\
&\Pr(s_t = k \mid { s_{t-1}=k_1, \ldots, s_{t-L} =k_L}, Z_t = r, \bm{z}_t = (\ell_1, \ldots, \ell_r), \bm{\mathcal{Q}}^{(r)}) \nonumber \\
& \quad = (\bm{\mathcal{Q}}^{(r)})_{ k, k_{\ell_1}, \ldots k_{\ell_r}} \, , 
\end{align}
where $\bm{\alpha}_{Q}$ is a length-$K$ vector of positive shape parameters (which could potentially be separately specified for each distribution in each $\bm{\mathcal{Q}}$); $\bm{\pi}_1$ and $\bm{\pi}_3$ are length-$R$ vectors containing probabilities such that $(\bm{\pi}_1)_{r} + (\bm{\pi}_3)_{r} < 1$; $\bm{\gamma}$ and $\bm{\delta}$ are length-$R$ vectors containing positive shape parameters; $\bm{\alpha}_{\lambda^{(r)}}$ is a length-$L \choose r$ vector of positive shape parameters; and $\beta_{\lambda^{(r)}} > 1$ is the SDM sparsity parameter. We always set $(\bm{\pi}_1)_{0} = 0$ to avoid penalizing $\Lambda_0$, and recommend setting $(\bm{\pi}_3)_{0} = 0$ as well. Note that all quantities in (\ref{eq:MMTDhier}) without explicit dependence are considered independent a priori.

As with the MTDg, posterior simulation can be accomplished entirely through closed-form Gibbs sampling. To simplify computation, we uniquely map all $Z_t$ and $\bm{z}_t$ pairs onto a single variable $\zeta_t \in \left\{ 0,1, \ldots, \left[{L \choose 1} + {L \choose 2} + \ldots + {L \choose R}\right]\right\}$ whose prior probability under the model is equal to the product of the corresponding $\Lambda$ and $\lambda$. Full conditional distributions for $\bm{\Lambda}$, each $\bm{\lambda}^{(r)}$, and each probability vector in $\{\bm{\mathcal{Q}}^{(r)}\}$ are exactly analogous to multinomial-SBM, multinomial-SDM, and multinomial-Dirichlet conjugate updates, respectively, where $Z_t$, $\bm{z}_t$, and observed data transitions supply the respective multinomial counts. Full conditional updates for $Z_t$ and $\bm{z}_t$ (equivalently $\zeta_t$) require calculation and sampling from a discrete distribution. Full details are given in Appendix \ref{sec:appendix_mmtd}.

We again improve mixing in the sampler by integrating $\{\bm{\mathcal{Q}}^{(r)}\}_{r=0}^R$ from the joint posterior, sampling the collapsed conditional distributions for $\bm{\Lambda}$, each $\bm{\lambda}^{(r)}$, and $\{\zeta_t\}$. These are supported by the tractable marginal distributions reported in Appendix \ref{sec:appendixA}. Additionally, to encourage occasional jumps between modes of the posterior, we include a hybrid independence-Metropolis step which jointly proposes $\bm{\Lambda}$, each $\bm{\lambda}^{(r)}$, and $\{\zeta_t\}$ from their joint prior every 10 iterations of the MCMC algorithm.

The augmented Gibbs sampler becomes computationally demanding as $R$ and $L$ increase because updates for the latent configuration variables $\{ \zeta_t \}$ involve calculation of $\sum_{r=0}^R {L \choose r}$ probabilities for each time point $t=L+1, \ldots, T$. 
Random-walk Metropolis samplers for $\bm{\Lambda}$, $\{\bm{\lambda}\}$, and $\{\bm{\mathcal{Q}}\}$ utilizing the mixture likelihood based on (\ref{eq:MMTDtrans}) may provide an alternate strategy if $K$ is reasonably small. 
The logit-normal distribution \citep{atchison1980}, or multivariate Gaussian random walks on the logit scale, facilitate properly constrained proposals for probability vectors.

\section{Simulation study}
  \label{sec:sims}

To demonstrate the effectiveness of the MMTD model for both objectives and to compare transition probability estimation performance with existing methods, we report two simulation studies. Both simulation scenarios feature time series generated from true Markov chains of differing order and lag configuration. In Simulation 1, the true generating model is a third-order chain with three states ($K=3$) in which transition probabilities depend on lags 1, 3, and 4. In Simulation 2, the true generating model is a fifth-order binary chain ($K=2$) for which each of the first five lags contributes to transition probabilities. In both models, each distribution in the transition tensor $\bm{\Omega}$ was drawn from a uniform distribution on the simplex (i.e., symmetric Dirichlet distributions with all shape parameters equal to 1). Each chain was randomly initialized and run for 1,000 steps of burn-in. The first 1,000 samples thereafter were reserved for training data and the next 1,000 for validation.

To evaluate estimation of transition probabilities, each model was fit using the prescribed number of training samples, and point estimates of the transition distributions were compared to the true transition distributions for each of the 1,000 validation points. Specifically, for validation time point $t'$, each model produced a vector $\hat{\bm{p}}_{t'}$ to estimate each $p_{t'}^{(k)} = \Pr(s_{t'} = k \mid s_{t'-1}, \ldots, s_{t'-L})= (\bm{\Omega})_{k, s_{t'-1}, \ldots, s_{t'-L}}$, for $k=1,\ldots,K$. In Bayesian models, the point estimate is the Monte Carlo-computed posterior mean of $\hat{\bm{p}}_{t'}$. In non-Bayesian models, $\hat{\bm{p}}_{t'}$ is computed from the optimized model fit. For each validation time point, we computed the $\mathcal{L}_1$ loss given by $L_{t'} = \sum_{k=1}^K \lvert \hat{p}_{t'}^{(k)} - p_{t'}^{(k)} \rvert$. The reported loss metric for model comparison is $100 \times \sum_{t'} L_{t'} / (KT')$, that is, 100 times the mean $\mathcal{L}_1$ loss across the $T' = 1{,}000$ validation points.

We fit the MTD, MTDg, and MMTD models to each training set with various settings. Implementation for the MTD is similar to the MTDg and is described in \citet{heiner2019spv}. Let $\text{MTD}(L)$ and $\text{MTDg}(L)$ denote the respective model fits with user-specified maximum lag horizon $L$, and let $\text{MMTD}(L,R)$ denote a model fit with user-specified maximum lag horizon $L$ and maximum order $R$. All transition distributions in all $\{\bm{Q}\}$ and $\{\bm{\mathcal{Q}}\}$ in all three models utilize symmetric, unit-information Dirichlet priors (i.e., whose shape parameters all equal $1/K$ so that they sum to unity).

We use two prior settings for the MTD model. The first employs a Dirichlet prior for $\bm{\lambda}$ with all shape parameters equal to $1/L$. The second setting uses a SBM prior for $\bm{\lambda}$ with $\pi_1 = 0.5$, $\pi_3=0.1$, $\eta=1{,}000$, and $\bm{\gamma}, \bm{\delta}$ selected to mimic the Dirichlet prior with shape parameters equal to $1/L$ and sparsity correction on $\bm{\delta}$ \citep{heiner2019spv}. This prior encourages a moderate level of sparsity as well as decreasing prior probability for higher lags.

The MTDg model uses a SBM prior for $\bm{\lambda}$ with $\pi_1 = 0$ for $\lambda_0$ and $\pi_1=0.5$ thereafter; $\pi_3=0$ for $\lambda_0$ and $\pi_3=0.2$ thereafter; $\eta=1{,}000$; $\gamma_0 = \delta_0 = 1$, yielding a uniform prior for $\lambda_0$, and remaining elements of $\bm{\gamma}$ and $\bm{\delta}$ selected to mimic a Dirichlet prior with shape parameters equal to $1/L$ and sparsity correction on $\bm{\delta}$. This prior avoids penalizing $\lambda_0$, encourages a moderate level of sparsity in the remaining lags, and steeper decrease in prior probability for higher lags than used for the MTD.

We use two prior settings in the MMTD models. Both follow (\ref{eq:MMTDhier}) with \\ 
 $\bm{\alpha}_{\lambda^{(r)}} = \left(1/{L \choose r}, \ldots, 1/{L \choose r}\right)$, but the first setting uses $\beta_{\lambda^{(r)}}=1$ for all $r=1,\ldots,R$, resulting in the symmetric, unit-information Dirichlet prior. The second setting uses $\beta_{\lambda^{(r)}}=\sqrt{T}$ to encourage selection of a single-lag configuration within level $r$. In both prior settings, we employ the SBM prior for $\bm{\Lambda}$ with $\pi_1 = 0$ for $\Lambda_0$ and $\pi_1=0.25$ thereafter; $\pi_3=0$ for $\lambda_0$ and $\pi_3=0.25$ thereafter; $\eta=1{,}000$; and $\gamma = \delta = 1$ for all second-component beta distributions, yielding a uniform prior for $\Lambda_0$. This prior avoids penalizing $\Lambda_0$, allows for sparsity in the remaining lags, and maintains soft ordering that favors lower levels of the model. Because $R$ is typically kept to small values, it is important that $\pi_1$ not be large and that $\pi_3$ not be too small. Otherwise, the prior can inappropriately allocate substantial mass toward large values of $\Lambda_R$. We recommend checking for this condition as part of prior sensitivity analysis.

To obtain results in Sections \ref{sec:sim1results}, \ref{sec:sim2results}, and \ref{sec:applications}, each model was initialized with random draws from Dirichlet distributions for $\bm{\Lambda}$ and each $\bm{\lambda}^{(r)}$. Random initialization in these models calls for long burn-in periods, on the order of tens to hundreds of thousands of iterations. In our analyses, 200,000 burn-in iterations were followed by another 400,000 iterations. Reported posterior quantities were calculated using a thinned sample retaining every 200th iteration. These conservative settings produced (unless otherwise noted) stable chains suitable for inference. In some cases, parallel chains sampled from MMTD models settled in neighboring modes which had minor impact on inferences and performance. With all models, posterior sampling was conducted using the {\em Julia} scientific computing language \citep{bezanson2017julia}.


In addition to our proposed models, we fit the multinomial generalized linear models with logistic link functions to each training set using the {\it VGAM} package in R \citep{yee2010vgam}. To distinguish different settings, we denote model fit as $\text{LogitMC}(L, R')$ with maximum lag horizon $L$ and highest interaction order among the linear predictors $R'$. We also fit the variable length Markov chain models, denoted VLMC, using the {\it VLMC} package in R \citep{vlmc} and employing default model settings.

 \subsection{Simulation 1 results}
  \label{sec:sim1results}

All models were fit to the time series from Simulation 1 for two sample sizes, $T=200$ and $T=500$. Here, we assume that the modeler is considering up to a horizon of six lags, which we use where possible to promote equitable comparisons. Results of the mean $\mathcal{L}_1$ loss across the 1,000 validation points are given in Table \ref{tab:sim1results}. In addition to transition probability estimation, we are interested in inferences for Markovian order and important lags afforded by the MTD, MTDg, and MMTD models. With exception of the MTD and MTDg, we see improved estimation with the larger sample size across all models.

\begin{table}[t]
	\centering
	\caption{Simulation 1 ($K=3$ states for a third-order chain with active lags 1, 3, and 4). Results for transition probability estimation under various models and model settings using two sample sizes, $T=200$ and $T=500$. The reported loss is 100 times the mean $\mathcal{L}_1$ loss, computed across 1,000 validation time points. Within each sample size group, the lowest mean loss is highlighted with bold font.}
	\label{tab:sim1results}
	\small
	
	\begin{tabular}{ccc}
		\begin{tabular}{l r}
			\toprule
			\multicolumn{2}{c}{$T=200$} \\
			Model & Loss \\
			\midrule
			LogitMC(6, 1) & 18.70 \\
			LogitMC(6, 2) & 37.42 \\
			LogitMC(3, 1), {\scriptsize Lags 1, 3, 4 only} & 17.14 \\
			LogitMC(3, 2), {\scriptsize Lags 1, 3, 4 only}  & 13.70 \\
			LogitMC(3, 3), {\scriptsize Lags 1, 3, 4 only}  & {\bf 12.64} \\
			&  \\
			VLMC & 19.12 \\
			& \\
			MTD(6), {\scriptsize Dir($\bm{\lambda}$)} & 17.29 \\
			MTD(6), {\scriptsize SBM($\bm{\lambda}$)} & 17.27 \\
			& \\
			MTDg(6) & 17.30 \\
			& \\
			MMTD(6, 2), {\scriptsize Dir($\bm{\lambda}$)} & 14.92 \\
			MMTD(6, 2) & 14.78 \\
			MMTD(6, 3), {\scriptsize Dir($\bm{\lambda}$)} & 14.65 \\
			MMTD(6, 3) & 13.72 \\
			MMTD(6, 4), {\scriptsize Dir($\bm{\lambda}$)} & 14.93 \\
			MMTD(6, 4) & 14.24 \\
			MMTD(6, 5), {\scriptsize Dir($\bm{\lambda}$)} & 15.13 \\
			MMTD(6, 5) & 14.40 \\
			\bottomrule
		\end{tabular}  
		&
		\hspace{0.1in}
		&
		\begin{tabular}{ l r }
			\toprule
			\multicolumn{2}{c}{$T=500$} \\
			Model & Loss \\
			\midrule
			LogitMC(6, 1) & 16.77 \\
			LogitMC(6, 2) & 18.84 \\
			LogitMC(3, 1), {\scriptsize Lags 1, 3, 4 only} & 16.39 \\
			LogitMC(3, 2), {\scriptsize Lags 1, 3, 4 only} & 10.40 \\
			LogitMC(3, 3), {\scriptsize Lags 1, 3, 4 only} & 7.64 \\
			&  \\
			VLMC & 15.26 \\
			& \\
			MTD(6), {\scriptsize Dir($\bm{\lambda}$)} & 17.27 \\
			MTD(6), {\scriptsize SBM($\bm{\lambda}$)} & 17.21 \\
			& \\
			MTDg(6) & 17.05 \\
			& \\
			MMTD(6, 2), {\scriptsize Dir($\bm{\lambda}$)} & 13.77 \\
			MMTD(6, 2) & 13.83 \\
			MMTD(6, 3), {\scriptsize Dir($\bm{\lambda}$)} & 7.55 \\
			MMTD(6, 3) & {\bf 7.44} \\
			MMTD(6, 4), {\scriptsize Dir($\bm{\lambda}$)} & 7.58 \\
			MMTD(6, 4) & 7.48 \\
			MMTD(6, 5), {\scriptsize Dir($\bm{\lambda}$)} & 7.56 \\
			MMTD(6, 5) & 7.47 \\
			\bottomrule

		\end{tabular}
		
	\end{tabular}
	
\end{table}

\subsubsection{Sample size 200}

In the $T=200$ case, the multinomial logistic models produce the best and worst results. Fitting all second-order interactions for up to six lags is cumbersome in this model, resulting in poor estimates. Fitting the full-order model to the correct lags only produces accurate estimates. However, this would require preliminary results from an iterative process which may or may not select the correct model and does not account for model uncertainty. We emphasize here that our proposed models do not require a model selection process if the modeler specifies the maximum lag horizon $L$ and maximum order $R$, as order and lag inferences are built-in.

The variable length Markov chain model offers no improvement, possibly because the dynamics governing Simulation 1 skip lag 2. VLMC branches utilizing more distant lags must pass through and include lag 2. This results in a missed opportunity for greater parsimony \citep{jaaskinen2014}.

The MTD and MTDg models offer little help in this scenario because Simulation 1 is third-order with non-additive interactions. Posterior densities for $\bm{\lambda}$ (not shown) reveal that $\lambda_3$ is favored under the Dirichlet prior (in the MTD) and dominates with the SBM prior (in both the MTD and MTDg). The latter effectively produces a first order Markov chain dependent on the third lag.

Several MMTD models were fit with increasing maximum order $R$ ranging from 2 to 5. The second-order model provides a substantial improvement over mixing first-order transitions and fits nearly as well as the correctly specified third-order model. As expected, estimation performance stops improving when $R$ exceeds the true order of three. Using SDM priors on $\bm{\lambda}$ parameters improves estimation, but more so when the correct model is contained in the specified model.

In the $R=2$ model, posterior inference supports second-order dynamics with the lag 3,4 combination receiving most posterior weight. Adding the SDM priors on $\{\bm{\lambda}^{(r)}\}$ results in stronger support for the same conclusion. In the $R=3$ model, posterior inferences support second or third-order dynamics with lags 3 and 4 receiving most posterior weight. Adding the SDM priors led to weakly favoring order 3 (selecting lags 1, 3 and 4) in one model run. The $R=4$ model most often supports second-order dynamics (lags 3 and 4) under both prior scenarios. Results from the $R=5$ model are similar to the $R=4$ model.

It appears that the signal associated with lag 1 is relatively weak when $T=200$. The SBM prior on $\bm{\Lambda}$ shrinks inferences toward second-order, but not decidedly away from third-order dynamics. Overall, the MMTD consistently produces the most faithful estimates of transition probabilities from a single model without requiring iterative model selection.

\subsubsection{Sample size 500}

In the $T=500$ case, even the multinomial logistic models fit directly to the correct lags only fail to outperform the MMTD with $R \ge 3$. With a larger sample size, the VLMC model is more competitive, but the MTD and MTDg remain insufficiently flexible to capture the structure. The MTD model mixes over lags 2, 4, and 5 with a Dirichlet prior on $\bm{\lambda}$, and primarily over lags 4 and 5 with the SDM prior. The MTDg concentrates some mass on lags 3 and 4.

In this large-sample scenario, MMTD performance improves substantially when the specified mixture model contains the true model structure (i.e., $R \ge 3$), although the second-order MMTD again improves over the first-order additive MTD and MTDg. In the MMTD models with $R \ge 3$, posterior mass concentrates on the correct order and lag configuration. Furthermore, we see little drop in performance when the maximal order is over-specified. 
The SDM priors on $\{\bm{\lambda}^{(r)}\}$ appear not to significantly improve estimation, and inferences are qualitatively similar. The inferences from the $R=2$ model are similar to those of the $R=2$ models fit to $T=200$ observations. Among the models considered in this simulation scenario, the MMTD consistently produces the most faithful estimates of transition probabilities.

 \subsection{Simulation 2 results}
\label{sec:sim2results}

All models were fit to the time series from Simulation 2 for three sample sizes: $T=100$, $T=200$ and $T=500$. Here, we assume that the modeler is considering up to a horizon of seven lags, which we use where possible to promote equitable comparisons. Results of the mean $\mathcal{L}_1$ loss across the 1,000 validation points are given in Table \ref{tab:sim2results}. Again, we examine order and lag inferences from the MTD, MTDg, and MMTD models in addition to estimation performance.

\begin{table}[t]
	\centering
	\caption{Simulation 2 ($K=2$ states for a fifth-order chain with five active lags). Results for transition probability estimation under various models and model settings using three sample sizes: $T=100$, $T=200$ and $T=500$. The reported loss is 100 times the mean $\mathcal{L}_1$ loss, computed across 1,000 validation time points. Within each sample size group, the lowest mean loss is highlighted with bold font.}
	\label{tab:sim2results}
	\small
	
	\begin{tabular}{ccccc}
		\begin{tabular}{l r}
			\toprule
			\multicolumn{2}{c}{$T=100$} \\
			Model & Loss \\
			\midrule
			LogitMC(7, 1) & 24.30 \\
			LogitMC(7, 2) & 26.15 \\
			LogitMC(7, 3) & n/a \\
			LogitMC(5, 1) & 24.49 \\
			LogitMC(5, 2) & 20.86 \\
			LogitMC(5, 3) & n/a \\
			LogitMC(5, 4) & n/a   \\
			&  \\
			VLMC & {\bf 20.52} \\
			& \\
			MTD(7)  & 24.71 \\
			\ \ \ {\scriptsize with $\Dirdist(\bm{\lambda})$} &  \\
			MTD(7) & 24.50 \\
			\ \ \ {\scriptsize with $\SBM(\bm{\lambda})$} &  \\
			& \\
			MTDg(7) & 24.01 \\
			& \\
			MMTD(7, 4)  & 23.68 \\
			\ \ \ {\scriptsize with $\Dirdist(\bm{\lambda})$} &  \\
			MMTD(7, 4) & 23.21 \\
			MMTD(7, 7)  & 23.68 \\
			\ \ \ {\scriptsize with $\Dirdist(\bm{\lambda})$} &  \\
			MMTD(7, 7) & 23.33 \\
			\bottomrule
		\end{tabular}  
		&
		\hspace{-0.05in}
		&
		\begin{tabular}{ l r }
			\toprule
			\multicolumn{2}{c}{$T=200$} \\
			Model & Loss \\
			\midrule
			LogitMC(7, 1) & 20.03 \\
			LogitMC(7, 2) & 16.26 \\
			LogitMC(7, 3) & 18.70 \\
			LogitMC(5, 1) & 20.25 \\
			LogitMC(5, 2) & 16.24 \\
			LogitMC(5, 3) & {\bf 11.26} \\
			LogitMC(5, 4) & n/a   \\
			&  \\
			VLMC & 15.45 \\
			& \\
			MTD(7) & 22.47 \\
			\ \ \ {\scriptsize with $\Dirdist(\bm{\lambda})$} &  \\
			MTD(7) & 23.31 \\
			\ \ \ {\scriptsize with $\SBM(\bm{\lambda})$} &  \\
			& \\
			MTDg(7) & 23.93 \\
			& \\
			MMTD(7, 4) & 15.26 \\
			\ \ \ {\scriptsize with $\Dirdist(\bm{\lambda})$} &  \\
			MMTD(7, 4) & 15.38 \\
			MMTD(7, 7) & 14.13 \\
			\ \ \ {\scriptsize with $\Dirdist(\bm{\lambda})$} &  \\
			MMTD(7, 7) & 13.93 \\
			\bottomrule
		\end{tabular}  
		&
		\hspace{-0.05in}
		&
		\begin{tabular}{ l r }
			\toprule
			\multicolumn{2}{c}{$T=500$} \\
			Model & Loss \\
			\midrule
			LogitMC(7, 1) & 18.53 \\
			LogitMC(7, 2) & 14.66 \\
			LogitMC(7, 3) & 13.67 \\
			LogitMC(5, 1) & 18.90 \\
			LogitMC(5, 2) & 15.35 \\
			LogitMC(5, 3) & 8.29 \\
			LogitMC(5, 4) & 7.79 \\
			&  \\
			VLMC & 12.13 \\
			& \\
			MTD(7) & 19.82 \\
			\ \ \ {\scriptsize with $\Dirdist(\bm{\lambda})$} &  \\
			MTD(7) & 19.59 \\
			\ \ \ {\scriptsize with $\SBM(\bm{\lambda})$} & \\
			& \\
			MTDg(7) & 19.68 \\
			& \\
			MMTD(7, 4) & 12.15 \\
			\ \ \ {\scriptsize with $\Dirdist(\bm{\lambda})$} & \\
			MMTD(7, 4) & 14.70 \\
			MMTD(7, 7) & 7.59 \\
			\ \ \ {\scriptsize with $\Dirdist(\bm{\lambda})$} & \\
			MMTD(7, 7) & {\bf 7.38} \\
			\bottomrule
		\end{tabular}  
		
	\end{tabular}
	
\end{table}

\subsubsection{Sample size 100}

In the $T=100$ case, high order interactions are not estimable in the multinomial logistic model. The VLMC model performs best in this scenario, presumably because Simulation 2 features no gap in relevant lags.

Because the simulation uses lags 1 through 5, the MTD(7), MTDg(7), and MMTD(7, 4) models are under-specified and must rely on a lower-order sub-model and/or mixing across lags to approximate the fifth-order dynamics. The MTD models mix primarily over lags 2 and 4, while the MTDg model concentrates on lag 2. The MMTD(7, 4) models mix primarily over low orders, with slight preference for lag 2. The over-specified MMTD(7, 7) models do not outperform the $R=4$ models, and produce qualitatively equivalent inferences for $\bm{\Lambda}$ and $\{ \bm{\lambda}^{(r)} \}$. It is apparent that the small sample size is insufficient to capture the fifth-order structure.

\subsubsection{Sample size 200}

With $T=200$, the time series is long enough to include third-order interactions in the multinomial logistic model, which performs well. The VLMC model is again competitive with the MMTD and generally outperforms the logistic models.

As before, the MTD and MTDg models are unable to leverage increased sample size to the extent that the other models can. The MTD models mix primarily over lags 1 and 5, while the MTDg model mixes primarily on lag 1 (due to the ordered prior on $\bm{\lambda}$). 

The higher-order interactions allowed by the MMTD become advantageous with $T=200$, making this model competitive. The MMTD(7, 4) models concentrate posterior mass on order four and lags 1, 2, 3, and 5. Posterior mass in the MMTD(7, 7) model is split between order four and five, again demonstrating the shrinking effect of the SBM prior on $\bm{\Lambda}$. Different runs of the MCMC chain favor lag configurations $(1,2,3,5)$ and $(1,2,3,4,5)$. SDM priors on $\{ \bm{\lambda}^{(r)} \}$ had a minor concentrating effect on the posterior densities.

We note that the over-specified MMTD(7, 7) with a less strictly ordered prior on $\bm{\Lambda}$ (such as the SDM) can outperform the correctly specified logistic model in this scenario. However, inferences from such models can be suspect, as they do not shrink toward the ``reduced'' and identifiable parameterization. While we favor reliably interpretable inferences, one may consider modifying or replacing the SBM prior on $\bm{\Lambda}$ to improve predictive performance.

\subsubsection{Sample size 500}

The fifth-order binary chain in Simulation 2 has 32 total (univariate) transition distributions which are easily estimated with 500 samples. Therefore, the multinomial logistic models with high-order interactions approach the performance of the over-specified MMTD models. The VLMC is also competitive. Again, the MTD(7) and MTDg(7) models lag noticeably behind in estimation performance, although both attempt to mix over multiple lags.

The MMTD(7, 4) with a Dirichlet prior on $\{ \bm{\lambda}^{(r)} \}$ again concentrates on order four and lags 1, 2, 3, and 5. The same model with the SDM prior selects lags 3, 4, 5, and 6, and performs noticeably worse (loss of 14.70). A second MCMC run places most weight on orders three and four, and lags 1, 3, 4, and 5, resulting in average $\mathcal{L}_1$ loss of 12.65. This highlights multimodality of the posterior and the need for replicate MCMC runs.

The MMTD(7, 7) decisively identifies the correct order and lag structure resulting in the best estimation performance. We conclude that the MMTD consistently produces the most faithful estimates of transition probabilities from a single model without requiring iterative model selection.

\section{Data illustrations}
 \label{sec:applications}

We now apply the MTDg and MMTD models to two data analyses. The first data example was studied with the original MTD and in the subsequent literature. The second is an analysis of pink salmon population dynamics in Alaska, U.S.A.\ during the twentieth century. We illustrate the use of inferences on order and lag importance available from the models.

\subsection{Seizure data}
\label{sec:seizure}

\citet{berchtold2002} demonstrate the MTD model using a binarized time series adapted from \citet{macdonald1997hmm}, which reports the occurrence of at least one epileptic seizure for a patient on each of 204 consecutive days. \citet{berchtold2002} fit several Markov chain and MTD models, using the Bayesian information criterion (BIC) to ultimately select a MTD with eight lags. They report that $\lambda_8$ has the greatest magnitude. Note that the MTD model used in \citet{berchtold2002} allows negative values in $\bm{\lambda}$, which requires a complex set of constraints for estimation. The seizure time series was revisited and fit using the methods in \citet{sarkar2016}, who report a model of maximal order 8, with lag 8 having the highest posterior inclusion probability. In contrast with \citet{berchtold2002}, they find lag 1 to be the second most important. They report the posterior mode for the number of important lags to be three.

In light of these two analyses, we fit the MTDg and MMTD to the seizure data with $L=10$ and $R=4$, each with prior settings identical to those used in the simulation studies. Trace plots (not shown) indicate that the marginal posterior distributions over $\bm{\Lambda}$ and each $\bm{\lambda}^{(r)}$ are multimodal, suggesting that more than one combination of lags could model the dynamics with similar accuracy. 
We note also that the assumption of time-homogeneity is questionable, as no seizures were reported in the last 29 days.

The MTDg model concentrates most posterior weight on lag 8, followed distantly by lags 4 and 9. The transition matrix for lag 8 suggests that the status eight days prior is most often replicated in the present (seizure or no seizure). This transition pattern is repeated for lags 4 and 9, producing a compounding effect for repeated seizures in the model, which effect we should emphasize is additive only. That this model clearly selects lag 8 demonstrates the utility of the SBM prior for the MTDg. The prior simultaneously shrinks parameters toward the identifiable model and maintains a conditional stochastic ordering on the lags while maintaining flexibility to select distant lags when this is supported by the data.

The MMTD(10, 4) model with Dirichlet priors on lag configurations mixes primarily over orders one and two. The standard model (with SDM priors on lag configuration weights) shifts more posterior weight to higher orders, with $\Lambda_2$ and $\Lambda_3$ edging one another in separate MCMC runs. Without clear selection of order and lags, we discourage over-interpretation of the transition probabilities in $\{ \bm{\mathcal{Q}}^{(r)} \}$. However, it is clear that the estimated probabilities favor persisting in previous states. For example, the posterior means for $(\bm{\mathcal{Q}}^{(2)})_{1,1,1}$ and $(\bm{\mathcal{Q}}^{(2)})_{2,2,2}$ are 0.78 and 0.73 respectively (posterior medians are 0.92 and 0.85). That is, no occurrence of seizure in recent days yields a high probability for no seizure on the current day, and repeated occurrence of seizures on multiple past days yields a high probability of seizure on the current day.


\begin{figure}[b!]
	\centering
	
	\begin{tabular}{c c} 
		\includegraphics[scale=0.60, trim = 5 0 13 0, clip]{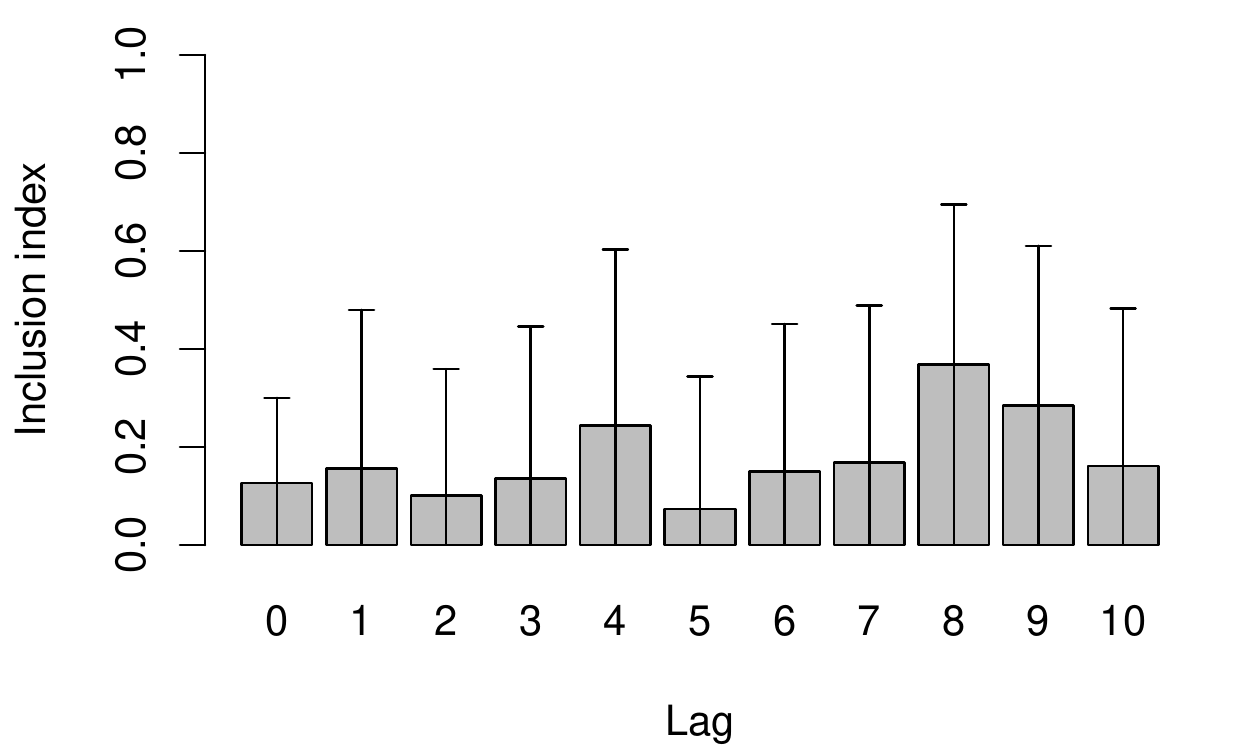} &
		\includegraphics[scale=0.60, trim = 5 0 13 0, clip]{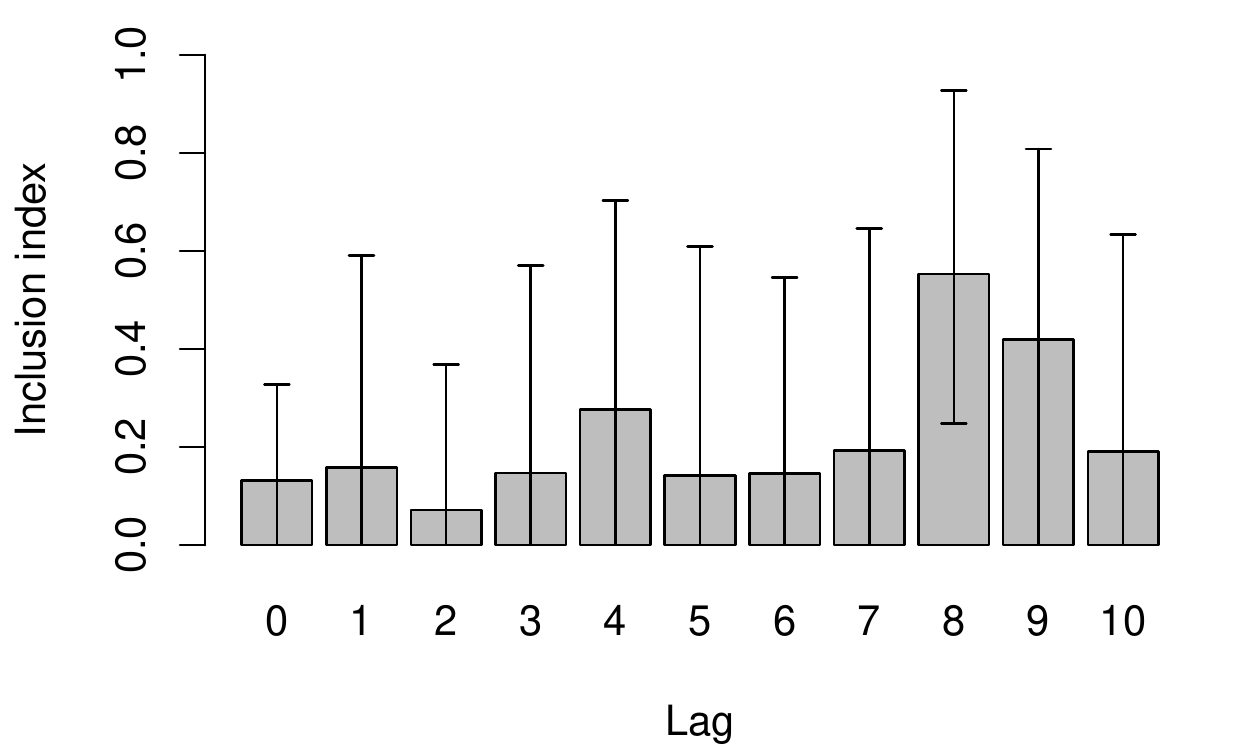}
	\end{tabular}
	
	\caption{ 
		Posterior mean (with 95\% credible interval) inclusion index for each lag in the seizure analysis, under the MMTD(10,4) model with the Dirichlet priors (left) and SDM priors (right) on lag configuration weights $\{ \bm{\lambda}^{(r)} \}$.
	}
	\label{fig:seizure_inclusion}
\end{figure}

We can more comprehensively assess lag importance by computing a lag inclusion index as the sum of all products $\Lambda_r \times \lambda_{(\bm{z}_j)}^{(r)}$ across $j=1,\ldots,{L \choose r}$ and $r=1,\ldots,R$ for which lag $\ell$ appears in the lag configuration $\bm{z}_j$. We compute this for each lag at each MCMC sample. Inference for $\Lambda_0$ is included as lag 0 for reference. Due to the shrinking SBM prior for $\bm{\Lambda}$, a high inclusion index for lag 0 should not be interpreted as a lack of Markovian dependence (unless it is near 1 with high confidence). However, a low inclusion index for lag 0 relative to other lags can indicate strong Markovian dependence. We summarize the inclusion index for the models fit to the seizure data in Figure \ref{fig:seizure_inclusion}, with bars reporting the posterior mean and whiskers reaching to the ends of 95\% posterior credible intervals. Note the large uncertainty for this inclusion index for all lags except lag 8 in the model with SDM priors on $\{ \bm{\lambda}^{(r)} \}$. We further note that the posterior inclusion pattern across lags (associated with the SDM priors for lag configuration weights) resembles a plot with similar interpretation in Figure 6 (e) of \citet{sarkar2016}. The most notable exception is that lag 1 does not feature prominently in our inferences. All analyses, including our three, agree that lag 8 is the most important in determining the transition probability.

\subsection{Pink salmon data}
\label{sec:salmon}

\begin{figure}[b!]
	\centering
	
	\begin{tabular}{cccc} 
		\multicolumn{4}{c}{\includegraphics[scale=0.6]{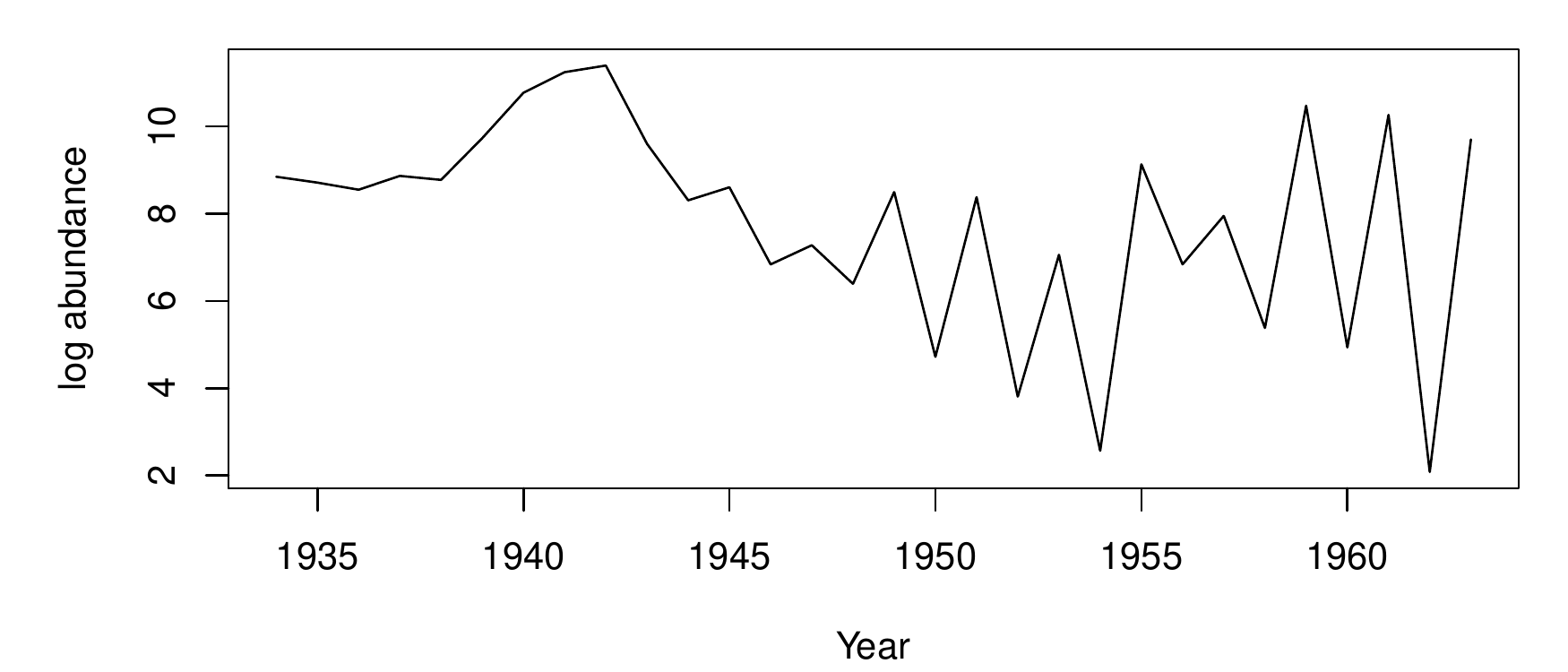}}
		\\
		\includegraphics[scale=0.45, page=2, trim = 10 3 15 1, clip]{plots/plots_pink_first30_discretized.pdf} & \hspace{-17pt}
		\includegraphics[scale=0.45, page=7, trim = 60 3 15 1, clip]{plots/plots_pink_first30_discretized.pdf} & \hspace{-17pt} \includegraphics[scale=0.45, page=12, trim = 60 3 15 1, clip]{plots/plots_pink_first30_discretized.pdf} & \hspace{-17pt}
		\includegraphics[scale=0.45, page=17, trim = 60 3 15 1, clip]{plots/plots_pink_first30_discretized.pdf} \\
	\end{tabular}
	
	\caption{Time-series plot (top) and bivariate lag plots (bottom) for the natural logarithm of pink salmon abundance from 1934 to 1963. In the lag plots, $y_t$ denotes abundance at time $t$ and horizontal/vertical lines separate $K=4$ quantile-based bins used to assign $\{y_t\}$ into discrete states $\{s_t\}$.}
	\label{fig:pinksalmon_timeseries}
\end{figure}

We next investigate a time series of annual pink salmon abundance (escapement) at a creek in Alaska, U.S.A.\ 
from 1934 to 1963 \citep{PinkSalmonData}. Population dynamics for pink salmon provide a testing opportunity for our model because pink salmon have a strict two-year life cycle \citep{heard1991pinksalmon}. Thus, we expect even lags to have the most influence in predicting the current year's population. A time-series plot of the natural logarithm of abundance is given in Figure \ref{fig:pinksalmon_timeseries} together with bivariate lag scatter plots. In this scenario, we might expect non-stationarity with long-term trends. It appears from the time series that the even-year population began to struggle in the late 1940s. Repeated interventions throughout the 1950s culminated in a population transfer in 1964 that bisects the complete time series and restricts us to the first segment \citep{PinkSalmonReport}. Nevertheless, the lag scatter plots suggest that we should be able to detect lag dependence structure, even with as few as 30 observations. After discretizing the data into sets of $K=4$ quantile-based bins using all 30 years, we fit the proposed models with the same prior settings used for the simulation studies. Because discretization is based on quantiles, results are invariant to monotonic transformations such as the natural logarithm.

\begin{figure}[b!]
	\centering
	
	\includegraphics[scale=0.44, page=1, trim = 2 10 20 7, clip]{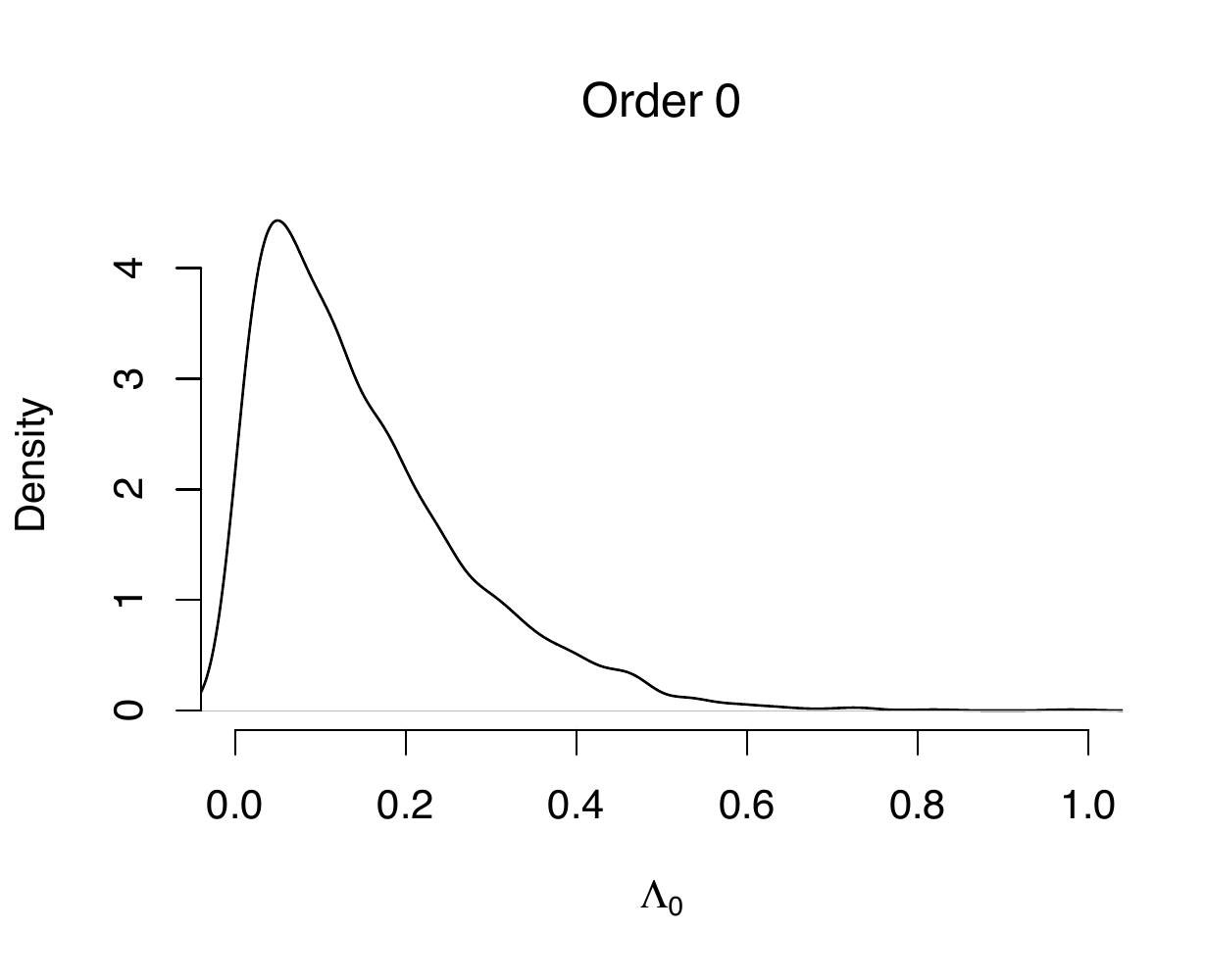} 	\includegraphics[scale=0.44, page=1, trim = 2 10 20 7, clip]{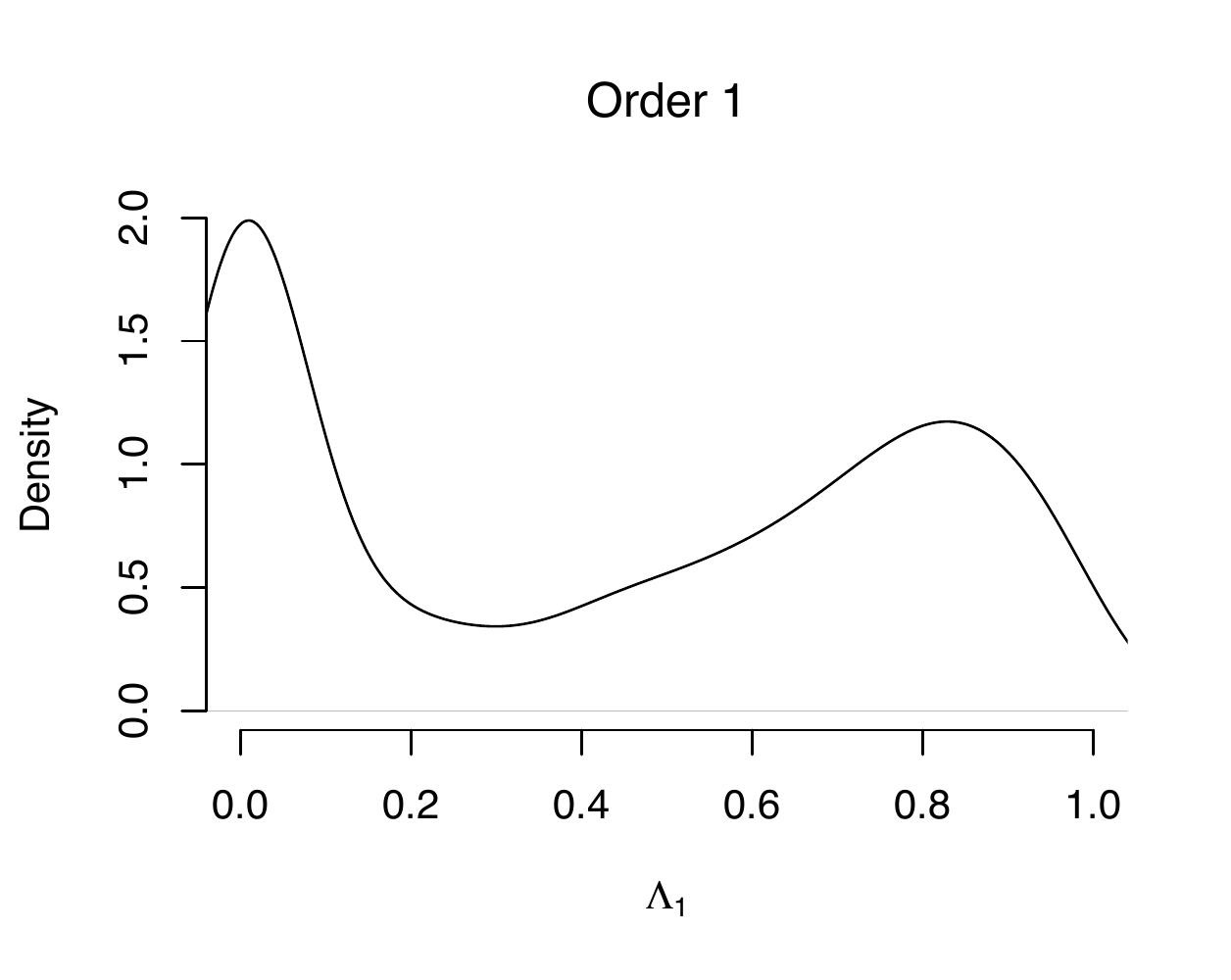} 	\includegraphics[scale=0.44, page=1, trim = 2 10 20 7, clip]{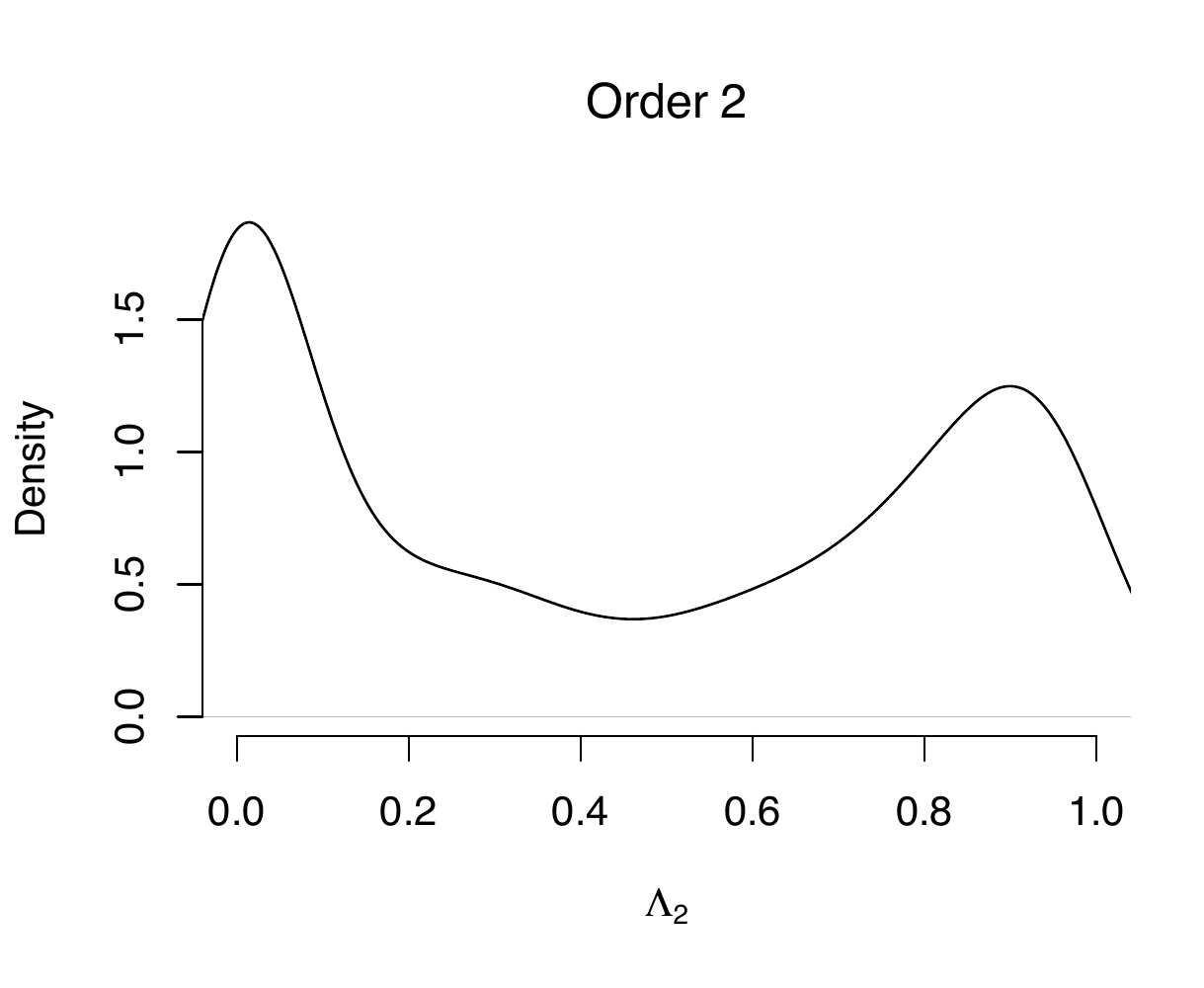}
	
	\caption{Marginal posterior density plots for $\bm{\Lambda}$ in the pink salmon analysis using a SBM prior on order and SDM priors for lag configuration weights.}
	\label{fig:pinksalmon_postdenslambda}
\end{figure}
\begin{figure}[t]
	\centering
	
	\includegraphics[scale=0.7]{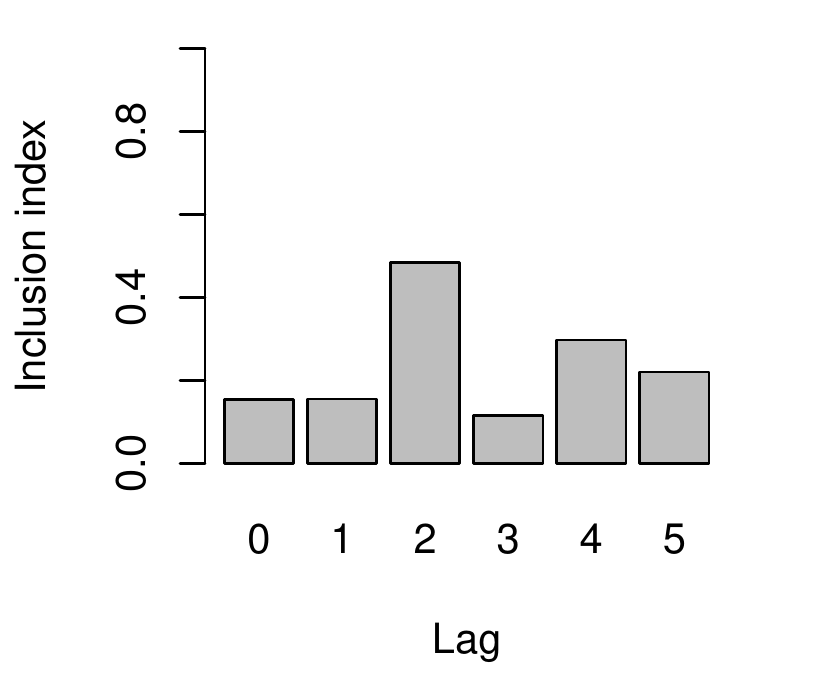}	\includegraphics[scale=0.7]{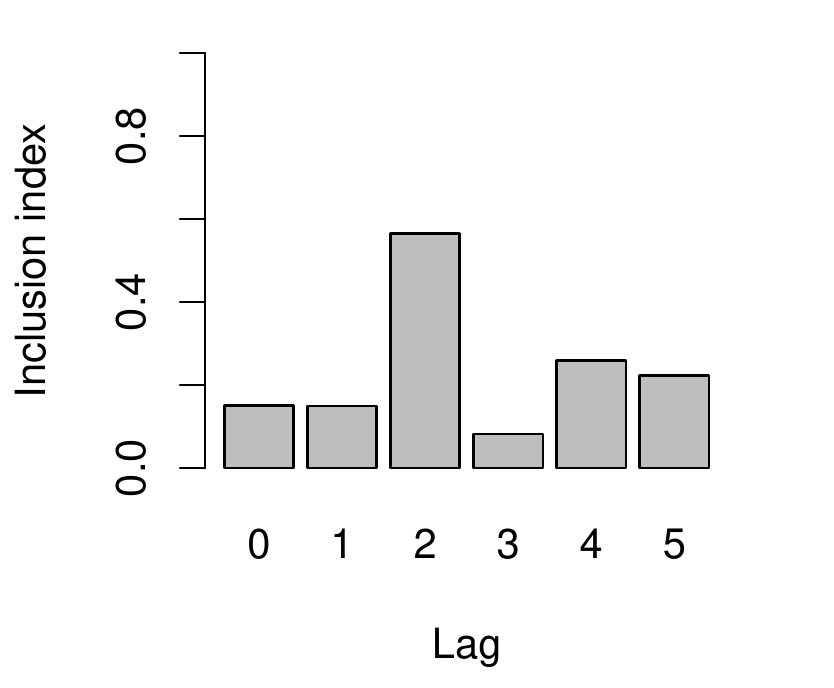}
	
	\caption{
		Posterior mean inclusion index for each lag in the pink salmon analysis under the MMTD(5,2) model with Dirichlet priors (left) and SDM priors (right) on lag configuration weights.}
	\label{fig:pinksalmon_laginclusion}
\end{figure}

The MTDg(5) model fit to the pink salmon time series clearly identifies lag 2 as the most influential (posterior means for $\lambda_\ell$, $\ell=0,1,\ldots,5$ are 0.17, 0.05, 0.67, 0.01, 0.07, and 0.03, respectively). The estimated $\bm{Q}^{(2)}$ also closely agrees with the lag-2 scatter plot in Figure \ref{fig:pinksalmon_timeseries}. In contrast, the MMTD(5,2) model shifts considerable posterior weight toward order two despite the shrinkage prior on order. Uncertainty, stemming from noisy dynamics and a small sample size, again results in a multimodal posterior, as seen in the density plots for $\bm{\Lambda}$ in Figure \ref{fig:pinksalmon_postdenslambda}. This uncertainty is also apparent in posterior inferences for the lag inclusion index. Credible intervals on the lag inclusion index are wide enough to warrant their omission from Figure \ref{fig:pinksalmon_laginclusion}. In these plots, we see essential agreement between the two prior settings for $\{ \bm{\lambda}^{(r)} \}$, with lag 2 being most prominent. Lags 4 and 5 also appear to contribute in some of the favored configurations.


\begin{figure}[t]
	\centering
	
	\includegraphics[scale=0.7, page=1, trim=0 0 0 35, clip]{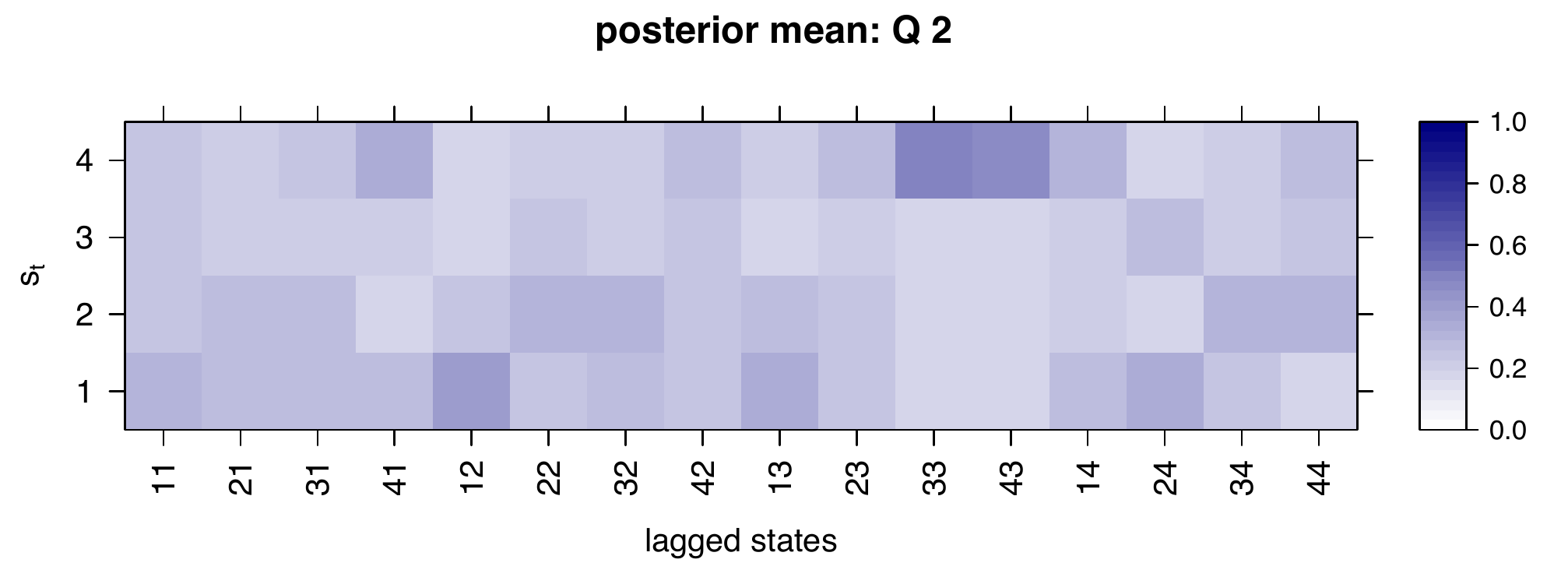}

	\caption{Posterior mean point estimate of the matricized $\bm{\mathcal{Q}}^{(2)}$ from the MMTD(5,2) pink salmon analysis with SDM priors on $\{ \bm{\lambda}^{(r)} \}$. Rows (along the $y$-axis) represent states to which the transition occurs, and columns (along the $x$-axis) represent the states occupied by the first two selected lags, with the state corresponding to the most recent lag changing index first.}
	\label{fig:pinksalmon_pmQ}
\end{figure}

It is important to examine the MMTD estimate of $\bm{\mathcal{Q}}^{(2)}$ to verify that the model is not attempting to fit first-order dynamics with a second-order chain. If this were the case, estimates of transition probabilities in $\bm{\mathcal{Q}}^{(2)}$ would repeat across the second lag index (in this case most likely representing lag 4 and/or 5). The posterior mean estimate of $\bm{\mathcal{Q}}^{(2)}$, shown in Figure \ref{fig:pinksalmon_pmQ}, appears not to have this problem, as consecutive $4 \times 4$ sub-matrices appear not to repeat. This is consistent under both priors. Hence, lag 2 may not be the only important lag.

\section{Summary}
 \label{sec:discussion}

We have explored two extensions of the original mixture transition distribution model for high-order Markov chains. The first is a Bayesian approach to a recent extension capable of identifying one or more important lags. The second captures higher-order interactions and can potentially yield useful inferences for order and lag importance. To accomplish the latter, the mixture of mixture transition distributions uses an over-specified model and sparsity-inducing priors to shrink back to an interpretable and informative structure. This is accomplished in a single model without necessitating iterative selection. Furthermore, our MCMC algorithm allows us to evaluate uncertainty about the model structure and transition probabilities. We demonstrated that this model can outperform some of the standard methods in transition probability estimation, and shown its practical utility in data analysis.


The over-specified MMTD model can offer insights into order and active lags provided the modeler approaches analysis attentively. In cases of large sample size or near-determinism, the true structure will immediately manifest in inferences for the mixture weight parameters. More often, lag importance should be aggregated and extracted in post-processing as we demonstrated in Section \ref{sec:seizure} with the lag inclusion index. If multiple lag patterns are prominent in the mixture model or different order components have non-overlapping lag patterns selected in each, the actual order of the time series may be higher than the highest selected model order. We also recommend checking the mixture transition tensors for redundancy, a sign of lower-order dependence. Absent a clearly identified lag structure through inferences on the mixture weights, we claim that this too can be informative.

Both the MTDg and MMTD models can approximate high-order dynamics by exploiting constructive additivity among lower-order transition probabilities. However, when this does not apply, a full jump to the next order in the MMTD is required. For example, in the salmon data analysis with five lags, the first-order mixture has five components associated with a $5\times 5$ transition matrix, whereas the second-order mixture has ten components associated with a $5 \times 5 \times 5$ transition tensor. A parsimonious compromise might rely on some factorization of the second-order tensor, as in \citet{sarkar2016}. We do not pursue this here, but rather choose to emphasize the interpretable structure of the proposed MMTD model (\ref{eq:MMTDtrans}) which showcases a model-averaging flavor with added flexibility.

\appendix

\section{Marginal distributions}
\label{sec:appendixA}

We report the marginal distributions of observations associated with the Dirichlet and SBM priors for probability vectors given originally in \citet{heiner2019spv}. These distributions can be useful for computing Bayes factors in addition to facilitating the MCMC algorithms described in Appendices \ref{sec:appendix_mtdg} and \ref{sec:appendix_mmtd}.

Consider a length-$N$ sequence of independent random variables $\{x_t\} \in \{1, \ldots, J\}^N$ with common distribution $\bm{\theta} = (\theta_1, \ldots, \theta_J)$. Given $\bm{\theta}$, the probability of the sequence is $\prod_{t} \theta_{x_t} = \theta_1^{n_1} \cdots \theta_K^{n_K}$ where the sufficient statistics in $\bm{n} = (n_1, \ldots, n_K)$ count the occurrences of each category.

If $\bm{\theta}$ follows a Dirichlet distribution with shape parameter vector $\bm{\alpha}$, then the marginal (prior predictive) distribution of $\{x_t\}$ is given by
\begin{align}
	\label{eq:DirMarg}
	p( \{x_t\} ) = \int p( \{x_t\} \mid \bm{\theta})\, p(\bm{\theta}) \, \diff \bm{\theta} = \frac{ \Gamma( \sum_j \alpha_j ) }{ \prod_j \Gamma(\alpha_j) }  \frac{ \prod_j \Gamma(\alpha_j + n_j) }{ \Gamma( \sum_j \alpha_j + n_j )  } 
	= \frac{ \MVBeta( \bm{\alpha} + \bm{n}) }{ \MVBeta( \bm{\alpha}) }\, ,
\end{align}
where $\MVBeta(\cdot)$ denotes the multivariate beta function. 

Now suppose $\bm{\theta}$ follows the SBM distribution with parameters $\{\pi_{1,j}\}$, $\{\pi_{3,j}\}$, $\eta$, $\{\gamma_j\}$, and $\{\delta_j\}$. Let 
\begin{align*}
	g_j(a_j, b_j, \bm{n}) \equiv \frac{\Gamma(a_j + b_j)}{\Gamma(a_j^* + b_j^*)} \frac{\Gamma(a_j^*)}{\Gamma(a_j)} \frac{\Gamma(b_j^*)}{\Gamma(b_j)} \, ,
\end{align*}
with $a_j^* \equiv a_j + n_j$, and $b_j^* \equiv b_j + \sum_{h=j+1}^K n_h \, $. Then the marginal distribution of $\{ x_t \}$ has probability mass function
\begin{align}
	\label{eq:SBMmarg}
	p(\{x_t\}) = \prod_{j=1}^{J-1} [ &\pi_{1,j} \, g_j(1, \eta, \bm{n}) + \pi_{2,j} \, g_j(\gamma_j, \delta_j, \bm{n}) + \pi_{3,j} \, g_j(\eta, 1, \bm{n}) ] \, .
\end{align}

\section{MCMC algorithm details: MTDg}
\label{sec:appendix_mtdg}

Following the hierarchical MTDg model outlined in (\ref{eq:hier_mtdg}), the joint posterior distribution of all unknown parameters is given up to proportionality:
\begin{align}
	\label{eq:fullJointPost_mtdg}
	p\big{(}\{z_t\}_{t=L+1}^T, \bm{\lambda}, \{\bm{Q}^{(\ell)}\}_{\ell=0}^L \mid \{s_t\}_{t=1}^T \big{)} &\propto p(\bm{\lambda}) \, p(\bm{Q}^{(0)}) \, \prod_{\ell=1}^L \prod_{k=1}^K \left[ p\left((\bm{Q}^{(\ell)})_{\cdot,k}\right) \right] \, \times  \\ 
	 &\quad \prod_{t=L+1}^T \left[ p(z_t \mid \bm{\lambda}) \, p\left(s_t \mid z_t, \{\bm{Q}^{(\ell)} \}_{\ell=0}^L , \{ s_{t-\ell} \}_{\ell=1}^{L} \right) \right]  \, , \nonumber
\end{align}
where $(\bm{Q}^{(\ell)})_{\cdot,k}$ denotes column $k$ from $\bm{Q}^{(\ell)}$.

\subsection{Full Gibbs sampler}
\label{sec:appendix_mtdg_original}

MCMC sampling for the full augmented model (\ref{eq:hier_mtdg}) can be achieved entirely with Gibbs updates. A Gibbs sampler cycles through the parameters, drawing updates from the conditional distributions given below.

\begin{itemize}
	\item $\Pr( z_t = \ell \mid \cdots ) \propto \Pr( z_t = \ell \mid \bm{\lambda} ) \, p\left(s_t \mid z_t, \{\bm{Q}^{(j)}\}_{j=0}^L , \{ s_{t-j} \}_{j=1}^L \right) \\ = \lambda_0 \, (\bm{Q}^{(0)})_{s_t} \, 1_{(\ell=0)} + \lambda_\ell \, (\bm{Q}^{(\ell)})_{s_{t}, s_{t-\ell}} \, 1_{(\ell \in \{1, \ldots, L \})}$, independently for each $t = L+1, \ldots, T $.
	\item $p(\bm{\lambda} \mid \cdots) \propto p(\bm{\lambda})\, \prod_t p(z_t \mid \bm{\lambda}) = \SBM( \bm{\lambda}; \bm{\pi}_1, \bm{\pi}_3, \eta, \bm{\gamma}, \bm{\delta}) \, \prod_t \lambda_{z_t} $, a conjugate SBM-multinomial update using the counts of $z_t$ in each of $\{0,1, \ldots, L\}$. A draw from the full conditional distribution begins by drawing the latent stick-breaking weights $X_\ell$ for $\ell=0,\ldots,L-1$, each from a mixture of three beta distributions. The mixture weights for $X_\ell$ are the three summands in the corresponding product terms of (\ref{eq:SBMmarg}) (with indexes shifted to begin at $\ell=0$), where $n_\ell$ is the cardinality of $\{t:z_t = \ell \}$. The three beta distributions have the corresponding $a_\ell^*$ and $b_\ell^*$ shape parameters taken from the SBM prior parameters and counts. The draw for $\bm{\lambda}$ is then constructed from the sampled $\{ X_\ell \}$ using (\ref{eq:stickbreaking}).
	\item $p(\bm{Q}^{(0)} \mid \cdots) \propto p(\bm{Q}^{(0)}) \prod_{t:z_t = 0} p\left(s_t \mid z_t, \{\bm{Q}^{(\ell)}\}_{\ell=0}^L, \{ s_{t-\ell} \}_{\ell=1}^L \right) \\ = \Dirdist( \bm{Q}^{(0)} ; \bm{\alpha}^{(0)} ) \prod_{t:z_t = 0} (\bm{Q}^{(0)})_{s_t} $, a standard conjugate Dirichlet-multinomial update using the counts of $s_t$ in each of $\{1, \ldots, K \}$ collected in $\bm{N}^{(0)}$. The full conditional is then $\Dirdist(\bm{\alpha}^{(0)} + \bm{N}^{(0)} )$.
	\item $p\left((\bm{Q}^{(\ell)})_{\cdot,k} \mid \cdots \right) \propto p\left((\bm{Q}^{(\ell)})_{\cdot,k}\right) \prod_{ \{t:z_t = \ell \text{ and } s_{t-\ell} = k \}} p\left(s_t \mid z_t, \{\bm{Q}^{(\ell)}\}_{\ell=0}^L, \{ s_{t-\ell} \}_{\ell=1}^L \right) \\ = \Dirdist \left( (\bm{Q}^{(\ell)})_{\cdot,k} ; \bm{\alpha}_k^{(\ell)} \right) \prod_{\{t:z_t = \ell \text{ and } s_{t-\ell} = k \}} (\bm{Q}^{(\ell)})_{s_t,k} $, a standard conjugate Dirichlet-\\multinomial update using the counts of $s_t$ in each of $\{1, \ldots, K \}$ collected in the $k$th column of $\bm{N}^{(\ell)}$. The full conditional is then $\Dirdist \left( \bm{\alpha}_k^{(\ell)} + (\bm{N}^{(\ell)})_{\cdot,k} \right)$, independent for each $\ell=1,\ldots,L$, and $k=1,\ldots,K$.
\end{itemize}

\subsection{Collapsed Gibbs sampler}
\label{sec:appendix_mtdg_modified}

Iterated full-conditional sampling of both $\{z_t \}$ and $\{\bm{Q}^{(r)}\}$ slows exploration of the joint posterior. To improve mixing, we instead integrate all $Q$ parameters out of (\ref{eq:fullJointPost_mtdg}) and conduct Gibbs sampling between $\{z_t \}$ and $\bm{\lambda}$. At each iteration, it is then straightforward to draw each $\bm{Q}^{(r)}$ from the conditional distributions given in Appendix \ref{sec:appendix_mtdg_original}.

Again we will summarize transition count information in $\{ (s_t, z_t) \}$ by aggregating into sufficient statistics $\bm{N}^{(0)}$ and $\{\bm{N}^{(\ell)}\}_{\ell=1}^L$, a set of $K \times K$ matrices described in Section \ref{sec:MTDg_inference} and Appendix \ref{sec:appendix_mtdg_original}. Integrating $\bm{Q}^{(0)}$ from (\ref{eq:fullJointPost_mtdg}) yields \\ $ p(\{z_t\}, \bm{\lambda} \mid \{s_t\}) \propto p(\bm{\lambda}) \, \prod_t \left[ p(z_t \mid \bm{\lambda}) \, p\left(s_t \mid z_t, \{ s_{t-\ell} \}_{\ell=1}^L  \right) \right] $ which differs from the original {\em only} in that
\begin{align}
	\label{eq:margLik}
	\prod_{t=L+1}^T \left[ p\left(s_t \mid z_t, \{ s_{t-\ell} \}_{\ell=1}^L \right) \right] = h(\bm{N}^{(0)}, \phi^{(0)}) \prod_{\ell=1}^L \prod_{k=1}^K h\left( (\bm{N}^{(\ell)})_{\cdot,k}, \phi_k^{(\ell)} \right) \, ,
\end{align}
where the $h(\cdot, \phi)$ takes the form of (\ref{eq:DirMarg}) if the columns of $\bm{Q}$ have independent Dirichlet priors, and (\ref{eq:SBMmarg}) if the columns of $\bm{Q}$ have independent SBM priors; and $\phi$ refers to generic hyperparameters appropriate for the choice of prior.

The modified algorithm then proceeds with the standard update for $\bm{\lambda}$ given in Appendix \ref{sec:appendix_mtdg_original}. Each $z_t$ is then updated individually with
\begin{align}
\label{eq:collapsedFC_mtdg}
\Pr( z_t = \ell \mid \bm{\lambda}, \{ s_t \}, \{ z_{t'} \}_{t' \ne t} ) \propto \lambda_\ell \, h(\bm{N}^{(0)}, \phi^{(0)}) \prod_{j=1}^L h\left( (\bm{N}^{(j)})_{\cdot,s_{t-\ell}}, \phi^{(j)}_{s_{t-\ell}} \right) \, ,
\end{align}
where the $\{\bm{N}^{(j)}\}$ are modified to reflect the possible values of $z_t \in \{0, 1, \ldots, L\}$.

\section{MCMC algorithm details: MMTD}
\label{sec:appendix_mmtd}

Following the hierarchical MMTD model outlined in (\ref{eq:MMTDhier}), the joint posterior distribution of all unknown parameters is given up to proportionality:
\begin{align}
\label{eq:fullJointPost}
p\big{(} \{ \zeta_t \}_{t=L+1}^{T}, & \bm{\Lambda}, \{ \bm{\lambda}^{(r)} \}_{r=1}^R, \{ \bm{\mathcal{Q}}^{(r)} \}_{r=0}^R,  \mid \{ s_t \}_{t=1}^T \big{)} \propto \nonumber \\
& \: p(\bm{\Lambda}) \, p(\bm{\mathcal{Q}}^{(0)}) \, \prod_{r=1}^R \left[ p( \bm{\lambda}^{(r)} ) \, \prod_{ j = 1 }^{ K^r} p\left((\bm{\mathcal{Q}}^{(r)})_{\cdot,j}\right) \right] \, \times  \\ 
& \prod_{t=L+1}^T \left[ p\left(\zeta_t \mid \bm{\Lambda}, \{ \bm{\lambda}^{(r)} \}_{r=1}^R \right) \, p\left(s_t \mid \zeta_t, \{\bm{\mathcal{Q}}^{(r)}\}_{r=0}^R, \{ s_{t-\ell} \}_{\ell=1}^L \right)  \right] \, , \nonumber
\end{align}
where $(\bm{\mathcal{Q}}^{(r)})_{\cdot,j}$ denotes column $j$ from a matricized version of $\bm{\mathcal{Q}}^{(r)}$.



\subsection{Full Gibbs sampler}
\label{sec:appendix_mmtd_original}

MCMC sampling for the full hierarchical model (\ref{eq:MMTDhier}) can be achieved entirely with Gibbs updates. A Gibbs sampler cycles through the parameters, drawing updates from the conditional distributions given below. In what follows, let $Z(\zeta)$ and $\bm{z}(\zeta)$ map $\zeta$ to its corresponding $Z$ and $\bm{z}$ respectively. Also, let $\varrho_r(\bm{s})$ be a unique map from each possible length-$r$ vector of lagged states $\bm{s} \in \{ 1, \ldots, K \}^r$ to the corresponding column index of the flattened (matricized) $\bm{\mathcal{Q}}^{(r)}$. Further, let $\bm{s}_{t-1}(\bm{z})$ be a function accepting a lag configuration $\bm{z}$ and returning the values of the states at those selected lags from the vector $(s_{t-1}, s_{t-2}, \ldots, s_{t-L})$. For example, if $\bm{z}_t = (2,5)$, then $\bm{s}_{t-1}(\bm{z}_t)$ will return the vector $(s_{t-2}, s_{t-5})$.

\begin{itemize}

	\item $\Pr( \zeta_t = i \mid \cdots ) \propto p\left(\zeta_t \mid \bm{\Lambda}, \{ \bm{\lambda}^{(r)} \}_{r=1}^R \right) \, p\left(s_t \mid \zeta_t, \{\bm{\mathcal{Q}}^{(r)}\}_{r=0}^R, \{ s_{t-\ell} \}_{\ell=1}^L \right) \\ \propto \Lambda_{Z(i)} \, \lambda_{ ( \bm{z}(i) ) }^{(Z(i))} \, (\bm{\mathcal{Q}}^{(Z(i))})_{s_t, \varrho_{Z(i)}( \bm{s}_{t-1}(\bm{z}(i)) )} $, independently for each $t \in \{L+1, \ldots, T\}$ with \\ $i \in \left\{0, 1, \ldots, \left[{L \choose 1} + {L \choose 2} + \ldots + {L \choose R}\right]\right\}$. Note that we define $\lambda^{(0)} \equiv 1$.

	\item $p(\bm{\Lambda} \mid \cdots) \propto p(\bm{\Lambda})\, \prod_t p(\zeta_t \mid \bm{\Lambda}, \{ \bm{\lambda}^{(r)} \}) \propto \SBM( \bm{\Lambda}; \bm{\pi}_1, \bm{\pi}_3, \eta, \bm{\gamma}, \bm{\delta}) \, \prod_t \Lambda_{Z(\zeta_t)} $, a conjugate SBM-multinomial update using the counts of $Z(\zeta_t)$ in each of $\{0,1, \ldots, R\}$.
	\item $p(\bm{\lambda}^{(r)} \mid \cdots) \propto p(\bm{\lambda}^{(r)})\, \prod_t p(\zeta_t \mid \bm{\Lambda}, \{ \bm{\lambda}^{(r)} \}) \propto \SDM(\bm{\lambda}^{(r)}; \bm{\alpha}_\lambda^{(r)}, \beta_\lambda^{(r)}) \,  \prod_{t : Z(\zeta_t) = r} {\lambda}_{(\bm{z}(\zeta_t))}^{(r)} $ independently for $r \in \{1, \ldots, R \}$. Here, $\bm{\lambda}^{(r)}$ is indexed by the $L \choose r$ possible sets of lags. This is a conjugate SDM-multinomial update using the counts of the ${L \choose r}$ unique lag configurations $\bm{z}_t$ within order $r$. The full conditional is a SDM distribution with $\beta_\lambda^{(r)}$ and with the multinomial counts added to $\bm{\alpha}_\lambda^{(r)}$, analogous to Dirichlet full conditionals.
	\item $p(\bm{\mathcal{Q}}^{(0)} \mid \cdots) \propto p(\bm{\mathcal{Q}}^{(0)}) \prod_{t:Z(\zeta_t) = 0} p\left(s_t \mid \zeta_t, \{\bm{\mathcal{Q}}^{(r)}\}_{r=0}^L , \{ s_{t-\ell} \}_{\ell=1}^L \right) \\ = \Dirdist( \bm{\mathcal{Q}}^{(0)} \mid \bm{\alpha}_{Q^{(0)}} ) \, \prod_{t:Z(\zeta_t) = 0} (\bm{\mathcal{Q}}^{(0)})_{s_t} $, a standard conjugate \\ Dirichlet-multinomial update using the counts of $s_t$ in each of $\{1, \ldots, K \}$ collected in $\bm{N}^{(0)}$. The full conditional is then $\Dirdist(\bm{\alpha}_{Q^{(0)}} + \bm{N}^{(0)} )$.
	\item $p\left( (\bm{\mathcal{Q}}^{(r)})_{\cdot,j} \mid \cdots \right) \propto p\left((\bm{\mathcal{Q}}^{(r)})_{\cdot,j}\right) \, \prod_{\{t : Z(\zeta_t)=r \text{ and } \varrho_r(\bm{s}_{t-1}(\bm{z}(\zeta_t))) = j \}} \, \times \\ p\left(s_t \mid \zeta_t, \{\bm{\mathcal{Q}}^{(r)}\}_{r=0}^R,\{ s_{t-\ell} \}_{\ell=1}^L \right) \\ \propto \Dirdist \left((\bm{\mathcal{Q}}^{(r)})_{\cdot,j} \mid \bm{\alpha}_{Q^{(r)}} \right) \, \prod_{\{t : Z(\zeta_t)=r \text{ and } \varrho_r(\bm{s}_{t-1}(\bm{z}(\zeta_t))) = j\}} (\bm{\mathcal{Q}}^{(r)})_{s_t,j} $, \\ independently for $r = 1, \ldots, R$, and $j = 1, \ldots, K^r $. Again, this is a standard conjugate Dirichlet-multinomial update using the transition counts collected in $(\bm{\mathcal{N}}^{(r)})_{\cdot, j}$, where $\bm{\mathcal{N}}^{(r)}$ is a matrix corresponding to the matricized version of $\bm{\mathcal{Q}}^{(r)}$. The full conditional distribution is then $\Dirdist(\bm{\alpha}_{Q^{(r)}} + (\bm{\mathcal{N}}^{(r)})_{\cdot, j} )$.
\end{itemize}

\subsection{Collapsed Gibbs sampler}
\label{sec:appendix_mmtd_modified}

Iterated full-conditional sampling of both $\{\zeta_t \}$ and $\{\bm{\mathcal{Q}}^{(r)}\}$ slows exploration of the joint posterior. To improve mixing, we instead integrate each $\bm{\mathcal{Q}}^{(r)}$ out of the joint posterior (\ref{eq:fullJointPost}) and conduct Gibbs sampling between $\{\zeta_t \}$, $\bm{\Lambda}$ and each $\bm{\lambda}^{(r)}$. At each iteration, it is then straightforward to draw each $\bm{\mathcal{Q}}^{(r)}$ from the conditional distributions given in Appendix \ref{sec:appendix_mmtd_original}.

For each $r = 1, \ldots, R$, again let $ \bm{\mathcal{N}}^{(r)} $ be a matrix containing transition counts for which the $(k, j)$ entry is the cardinality of $\{t : Z(\zeta_t)=r \ \text{and} \ \varrho_r(\bm{s}_{t-1}(\bm{z}(\zeta_t))) = j \ \text{and} \ s_t = k \}$. Also let the $k$th entry of vector $\bm{N}^{(0)}$ be the cardinality of $\{ t: Z(\zeta_t)=0 \ \text{and} \ s_t = k \}$. Integrating all $\bm{\mathcal{Q}}^{(r)}$ from the full joint posterior proportional to (\ref{eq:fullJointPost}) yields 
\begin{align}
\label{eq:collapsedJointPost}	
	p\left( \{ \zeta_t \}, \bm{\Lambda}, \{ \bm{\lambda}^{(r)} \} \mid \{ s_t \} \right) & \propto \SBM(\bm{\Lambda}) \, \prod_r \left[ \SDM( \bm{\lambda}^{(r)} ) \right] \, \prod_t \left[ \Lambda_{Z(\zeta_t)} \, \lambda_{ ( \bm{z}(\zeta_t) ) }^{(Z(\zeta_t))} \right] \times \nonumber \\ 
	& \quad p\left( \bm{N}^{(0)} \mid \{ \zeta_t \}, \{ s_t \} \right) \prod_{r=1}^R \left[ \prod_{j=1}^{K^r} p\left( (\bm{\mathcal{N}}^{(r)})_{\cdot,j} \mid \{ \zeta_t \}, \{ s_t \} \right) \right] \, ,
\end{align}
where $ p\left( (\bm{\mathcal{N}}^{(r)})_{\cdot,j} \mid \{ \zeta_t \}, \{ s_t \} \right) $ takes the form of (\ref{eq:DirMarg}) if the columns of matricized $\bm{\mathcal{Q}}^{(r)}$ follow independent Dirichlet priors, 
and (\ref{eq:SBMmarg}) if they follow independent SBM priors. The same marginal distribution forms apply for $\bm{N}^{(0)}$.

The modified algorithm then proceeds with the standard updates for $\bm{\Lambda}$ and each $\bm{\lambda}^{(r)}$ given in Appendix \ref{sec:appendix_mmtd_original}. Each $\zeta_t $ is then updated individually using its discrete collapsed conditional 
\begin{align}
	\label{eq:collapsedConditionalZeta}	
	p\left( \zeta_t \mid \cdots, -\{ \bm{\mathcal{Q}}^{(r)} \} \right) &\propto \Lambda_{Z(\zeta_t)} \, \lambda_{ ( \bm{z}(\zeta_t) )}^{(Z(\zeta_t))} p\left( \bm{N}^{(0)} \mid \{ \zeta_t \}, \{ s_t \} \right) \, \times \nonumber \\ 
	& \quad  \prod_{r=1}^R \left[ \prod_{j=1}^{K^r} p\left( (\bm{\mathcal{N}}^{(r)})_{\cdot,j} \mid \{ \zeta_t \}, \{ s_t \} \right)\right] \, ,
\end{align}
where we modify $\{ \bm{\mathcal{N}}^{(r)}  \} $ to reflect each possible value of \\ 
$\zeta_t \in \left\{0, 1, \ldots, \left[{L \choose 1} + {L \choose 2} + \ldots + {L \choose R}\right]\right\}$.


	
	
	
	


\bibliographystyle{jasa3}

\bibliography{references}

\end{document}